\documentclass[aps, pra, letterpaper, 10pt, twocolumn]{revtex4-2}
\usepackage[colorlinks=true,allcolors=blue!80!black]{hyperref}
\usepackage{amsmath, amsfonts, amssymb}
\usepackage{mathptmx}
\usepackage[mathscr]{euscript}
\usepackage{graphicx, epsfig, subfigure}
\usepackage{color, xcolor}
\usepackage{bm, bbm}
\usepackage{mathtools}
\usepackage{xfrac}
\DeclareMathAlphabet{\mathcal}{OMS}{cmsy}{m}{n}
\definecolor{physreva}{RGB}{097, 109, 112}
\definecolor{physrevb}{RGB}{186, 010, 045}
\definecolor{physrevc}{RGB}{000, 115, 107}
\definecolor{physrevd}{RGB}{053, 115, 133}
\definecolor{physreve}{RGB}{141, 085, 052}
\definecolor{physrevx}{RGB}{000, 077, 171}
\definecolor{physrevlett}{RGB}{66, 115, 92}

\newcommand{\ket}[1]{\vert{#1}\rangle} 
\newcommand{\bra}[1]{\langle{#1}\vert}

\newcommand{\mean}[1]{\langle #1 \rangle}

\newcommand{\tr}[2]{\mathrm{Tr}_{#1}\!\left\{#2\right\}}
\newcommand{\hatd}[1]{\hat{#1}^{\dagger}}

\renewcommand\Re{\operatorname{Re }}

\newcommand{\half}{\tfrac{1}{2}}
\newcommand{\one}{\openone}
\newcommand{\sighat}{\hat{\sigma}}

\newcommand{\Nbos}{\overline{n}}

\begin{document}
\title{Continuous variable multipartite vibrational entanglement}

\author{Mehdi Abdi}
\email{mehabdi@gmail.com}
\affiliation{Department of Physics, Isfahan University of Technology, Isfahan 84156-83111, Iran}

\begin{abstract}
A compact scheme for the preparation of macroscopic multipartite entanglement is proposed and analyzed. 
In this scheme the vibrational modes of a mechanical resonator constitute continuous variable (CV) subsystems that entangle to each other as a result of their interaction with a two-level system (TLS).
By properly driving the TLS, we show that a selected set of modes can be activated and prepared in a multipartite entangled state.
We first study entanglement properties of a three-mode system by evaluating the genuine multipartite entanglement.
And investigate its usefulness as a quantum resource by computing the quantum Fisher information.
Moreover, the robustness of the state against the qubit and thermal noises is studied, proving a long-lived entanglement.
To examine the scalability and structural properties of the scheme, we derive an effective model for the multimode system through elimination of the TLS dynamics.
This work provides a step towards a compact and versatile device for creating multipartite noise-resilient entangled state in vibrational modes as a resource for CV quantum metrology, quantum communication, and quantum computation.
\end{abstract}
%\date{\today}
\maketitle

%
%
%========================================%
\section{Introduction}%
Quantum entanglement is a pivotal resource in many quantum technologies, from quantum communications~\cite{Bennett1993, Braunstein1998, Kimble2008} and quantum computations~\cite{DiVincenzo1995, Lloyd1999, Menicucci2006}, to quantum metrology where the Heisenberg limit is approached only by employing a multipartite entangled state~\cite{Wineland1992, Huelga1997, Giovannetti2011}.
Meanwhile, as mechanical systems have proven to be very efficient in weak force sensing both at macroscopic level~\cite{Binnig1986, Chan2001} and mesoscopic scales~\cite{Gavartin2012, Moser2013}, the quantum sensing thus would require investigations on the entangled mechanical oscillators~\cite{Eisert2004, Plenio2004, Pirandola2006, Jost2009, Abdi2012, Ockeloen2018, Riedinger2018}.
From a different perspective, the entangled mechanical resonators can serve as nodes of a quantum network~\cite{Schmidt2012, Rips2013, Zippilli2021}.
%The entanglement of mechanical resonators to the optical field of a cavity is a parallel route towards the manipulation of mechanical resonators~\cite{} presenting them as an alternative platform for quantum computation (QC).

%For universal quantum computation (QC) with continuous variable (CV) systems two different paradigms can be envisioned as with the discrete variables: The quantum circuit model where the quantum modes (qumodes) are initialized, operated, and readout.
%And the measurement-based approach where a network of qumodes are prepared in an entangled quantum state as a resource~\cite{Bremner2009} on which the computations are performed by measurements.
%The challenge in the latter is to create a cluster state in which many modes are entangled to each other.
%The measurements only need to be local, making them significantly feasible.
%The universality of the QC by a cluster state is achieved by, first, availability of a non-Gaussian feature~\cite{Menicucci2006}, and second, forming a tight enough network, i.e. a square lattice~\cite{Raussendorf2001}.
%The non-Gaussianity of the whole cluster is one option that is pursued.
%Alternatively, one could work with non-Gaussian measurements~\cite{Weedbrook2012}.

Interaction of the mechanical resonators with other quantum systems such as optical cavities~\cite{Vitali2007, Paternostro2007, Favero2009}, electronic~\cite{Blencowe2004,Abdi2015a}, and spin degrees of freedom~\cite{Lee2017} opens an avenue for employing their vibrational modes for quantum information processing~\cite{Hartmann2008, Rabl2010b, Stannigel2010, Abdi2015b}.
An opposite route is to use the mechanical resonators for controlling~\cite{MacQuarrie2013, Barfuss2015, Schuetz2015, Maity2020} or intermediating interactions among other physical systems~\cite{Bennett2013, Albrecht2013}.
Such schemes have already been sought and used for developing scalable quantum networks~\cite{Lemonde2018, Safavi2019}.
In either case, the size of the system can be considerably miniaturized---compared to the electromagnetic counterparts---due to the smaller wavelength and stationary nature of the vibrational modes.
Multimode optomechanical systems have been investigated from various points of view both theoretically~\cite{Xuereb2012, Xu2013, Seok2013} and experimentally~\cite{Shkarin2014, Fan2015, Nielsen2017, Arrangoiz2019}.
Recently the multimode circuit quantum acousto-dynamical (cQAD) systems based on bulk~\cite{Han2016, Kervinen2019, Gokhale2020} and surface~\cite{Manenti2017, Moores2018, Satzinger2018, Sletten2019, Bienfait2019} acoustic resonators have been subject to an intense research, propelling towards the scalable quantum information with vibrational modes~\cite{Houhou2015, Moore2017, Tan2017, Houhou2018, Hann2019}.

\begin{figure}[tb]
\includegraphics[width=\columnwidth]{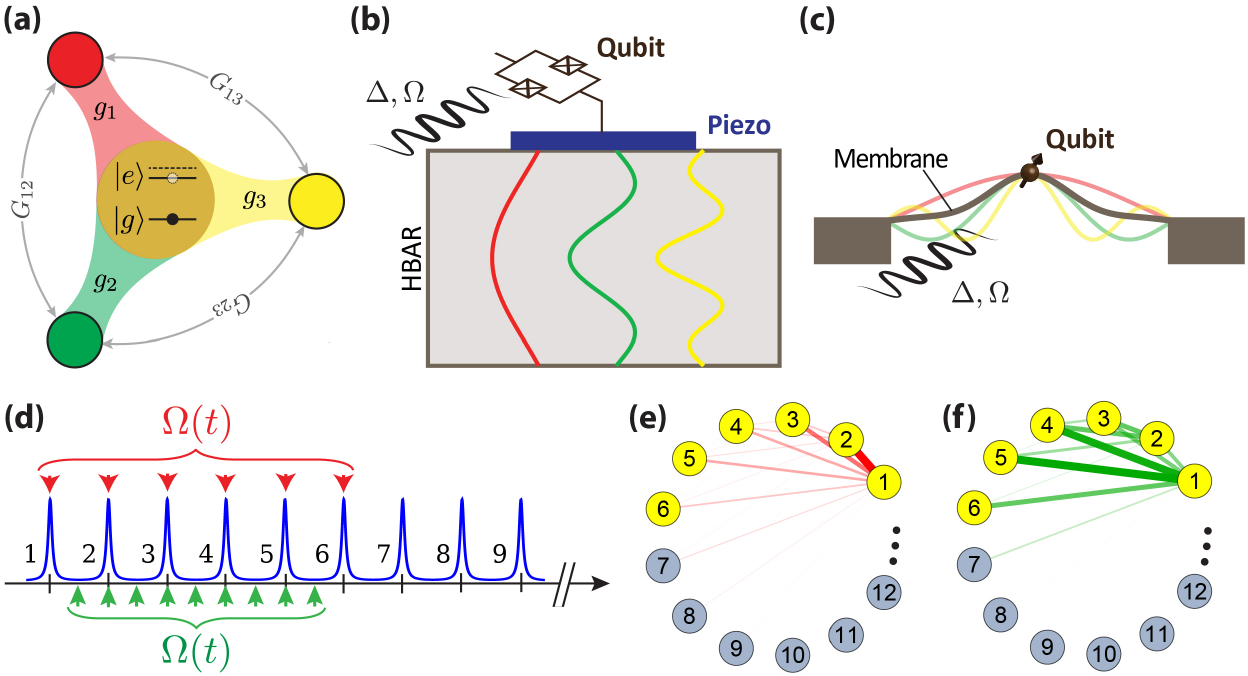}
\caption{%
(a) A TLS coupled to multiple vibrational modes of a mechanical resonator can intermediate their mutual interactions, which allows for creating a multipartite entangled state among those modes. This can be implemented e.g. in:
(b) A HBAR whose longitudinal modes couple to a superconducting qubit via a piezoelectric device.
Or (c) the flexural modes of a membrane interacting with an embedded atom-like defect.
(d) In a multimode scheme the set of modes are \textit{activated} (here 1 to 6) by employing a properly modulated drives that form two-mode squeezing interactions.
The interaction is shown as a graph for two different modulation schemes: At the mode frequencies (e) and the half sum frequencies (f). The edge thickness represents the interaction strength, while the \textit{active} modes are highlighted in yellow.}
\label{fig:scheme}%
\end{figure}

In this work, we put forth a scheme that allows for creating multipartite vibrational entangled states in a controllable way via their coupling to a two-level system (TLS).
Instead of several mechanical systems, we propose to employ the normal modes of a single resonator.
Thus, significantly reducing spatial extent, complexity, and noise of the system.
The TLS serves in mediating a controllable interaction between the modes.
The scheme is in principle implementable in surface acoustic wave (SAW) cavities and high overtone bulk acoustic resonators (HBAR) coupled to superconducting qubits in a cQAD system~\cite{Chu2017, Andersson2020} or can be envisaged in other setups such as flexural modes of a membrane combined with an embedded optically active lattice defect~\cite{Abdi2018a, Abdi2018b}.
Here, we study a regime where the TLS decay rate dominates its coupling to the vibrational modes and show that the TLS can be adiabatically eliminated from the system dynamics to obtain a network of effectively interacting modes with an adjustable interaction structure.
The latter is achieved by modulating drive frequency of the TLS at the eigenfrequencies of a set of vibrational modes, see Fig.~\ref{fig:scheme}.
We show that a multipartite entangled state of CV systems is attainable under experimentally realizable conditions and study its robustness against the system noises.
%Thanks to the nonlinear essence of the TLS, the resulting GHZ-state is non-Gaussian.
Furthermore, by computing the quantum Fisher information (QFI) of the state we prove that these states are useful for enhanced quantum sensing at the Heisenberg limit, e.g. for detection of gravitational waves and dark matter evidences~\cite{Goryachev2014, Arvanitaki2016, Abdi2016}.
%The structure of the entangled state and their robustness against the noise and imperfections is also studied.
%
%

%
%
%========================================%
\section{Model}%
The frequency spectrum $\{\omega_k\}$ ($k=1,2,\cdots, N$) of the vibrational eigenmodes of a mechanical resonator are determined by its geometrical properties~\cite{Renninger2018, Abdi2018b}.
These vibrational degrees-of-freedom can couple to a two-level system in various schemes, e.g. a superconducting qubit through piezoelectric effects in a cQAD device~\cite{Chu2017} or to the electronic spin two-level system of an embedded optically active lattice defect~\cite{Kolkowitz2012, Muschik2014, Li2016, Abdi2017, Abdi2019}. 
The dynamics of the system is governed by the Hamiltonian
\begin{equation}
\hat{H} = \half(\Delta\sighat_z +\Omega\sighat_x) +\sum_{k=1}^{N}[\omega_k\hatd{b}_k\hat{b}_k + \half g_k\sighat_z (\hat{b}_k +\hatd{b}_k)],
\label{ham}
\end{equation}
where $\sighat_x$ and $\sighat_z$ are the Pauli matrices of a TLS driven with a Rabi frequency $\Omega$ and at detuning $\Delta$.
The vibrational modes are expressed by the phonon annihilation (creation) operators $\hat{b}_k$ ($\hatd{b}_k$) with the nontrivial canonical commutation relation $[\hat{b}^{}_k,\hatd{b}_l]=\delta_{kl}$.
The dimensionless canonical operators of each mode are related to these bosonic operators through $\hat{b}_k = (\hat{x}_k +i\hat{p}_k)/\sqrt{2}$
% the zero-point motion amplitude $x_{\mathrm{zp},k}$ as $\hat{X}_k=\xzpm(\hat{b}_k^{} +\hatd{b}_k)$ and the conjugate momentum $\hat{P}_k$
such that $[\hat{x}_k,\hat{p}_l]=i\delta_{kl}$.

Before moving to the full analysis of the system, we qualitatively discuss how the Hamiltonian in Eq.~\eqref{ham} leads to the entanglement of vibrational modes.
Interaction of the TLS with the spectrum of the mechanical modes leads to their mutual interactions [Fig.~\ref{fig:scheme}(a)].
This can be readily seen by adiabatic elimination of the TLS from the equations of motion.
The process is valid when the TLS decay rate $\Gamma$ is greater than its coupling strength to any mechanical mode of interest, for details see Appendix~\ref{sec:adi}.
We also assume that the TLS is driven at far off-resonance ($\Delta \gg \Gamma,\Omega$) and apply a proper polaron transformation~\cite{Rabl2010a}.
The resulting interaction Hamiltonian is not in resonance with the desired interactions, and thus, does not provide appreciable entanglements.
To `activate' the interactions, we propose to modulate the TLS drive $\Omega(t)=\Omega_0\sum_i\cos(w_i\hspace{0.5mm}t)$ at the proper frequencies $w_j$~\cite{Galve2010, Abdi2015b}.
This brings us at the following interaction Hamiltonian after applying a rotating wave approximation (RWA)
\begin{equation}
\hat{H}_{\rm RWA} = \half\sum_{k,l} G_{k,l}\big(B_{k,l}^{\rm tms}\hat{b}_k\hat{b}_l -B_{k,l}^{\rm qst}\hat{b}_k\hatd{b}_l\big) +\mathrm{H.c.},
\label{eff}
\end{equation}
where $G_{k,l}$ is the effective coupling rate and $B^{\rm tms}$ and $B^{\rm qst}$ are the weighted adjacency matrices that determine the strength of two-mode squeezing and quantum state transfer interactions, respectively, see Appendix~\ref{sec:rwa} for the details.
%These can be visualized as graphs with weighted edges.
When the mechanical spectrum has non-commensurate frequencies, any desired subset of modes are activated by setting $\{w_i\}\subseteq \{\omega_k\}$ with the cardinality $\text{card}\{w_i\}=M$, where $M$ is the number of \textit{active} modes.
In the language of graphs, the system forms a complete-graph.
Such interactions with enough strength lead to a state which is equivalent to a multipartite CV GHZ-state~\cite{Pfister2004, Bradley2005, Zhang2006, Briegel2009}.
In a commensurate spectrum, the other modes get involved in the interactions by the same choice of modulation frequencies.
However, their coupling strength is smaller than the \textit{active} modes and only negligibly contribute in the entanglement dynamics [Fig.~\ref{fig:scheme}(e)].
Alternatively, one modulates the drive at half of the mode sum frequencies $w_i=\half(\omega_k+\omega_l)$ for getting a better connected graph [Fig.~\ref{fig:scheme}(f)]. Nonetheless, the numerical results suggest that the effect is incremental at the cost of a more complicated modulation.
% as the number of required frequencies are $\binom{M}{2}$, whilst in the former case it equals the number of active modes $M$.

In a HBAR cQAD device the longitudinal modes of a bulk acoustic wave cavity with high quality factors couple to a superconducting qubit through a piezoelectric interface, see e.g.~\cite{Gokhale2020}. The HBAR modes form a spectrum of equally spaced frequencies $\omega_k=k\delta_{\rm FSR}$ ($k=1,2,\cdots$) with free spectral range $\delta_{\rm FSR}$ which is determined by the cavity length and medium, hence, forming a commensurate spectrum.
A few techniques can be envisaged for introducing anharmonicity to the mode spacing~\cite{Hann2019}, and thus, enhancing efficiency of the activating protocol.
However, here we set our focus on the simple commensurate setup for experimental feasibility and concreteness.
%Alternatively, one could consider a membrane under a homogenous dominant tensile strain with embedded color centers for implementing the idea~\cite{Muschik2014, Abdi2019, Abdi2019b}.

%
%
%========================================%
\section{Triangle system}%
%We first study three active modes.
The full system is described by Hamiltonian \eqref{ham} where the nonlinear essence of the qubit gives rise to non-Gaussianity of the mechanical state.
%A feature that is of vital importance in universal quantum computation.
As the simplest network, first we investigate a triangle; a system made of three lowest vibrational modes [Fig.~\ref{fig:scheme}].
The dynamics is described by the quantum optical master equation
\begin{align}
\label{master}
\dot\rho = \frac{i}{\hbar}\big[\rho,\hat{H}\big]
&+\frac{\Gamma}{2}\big\{\!(\Nbos_{\omega_q}+1)\mathcal{D}_{\sighat_-}\![\rho] +\Nbos_{\omega_q}\mathcal{D}_{\sighat_+}\![\rho] \!\big\} +\frac{\widetilde\Gamma}{2}\mathcal{D}_{\sighat_z}[\rho] \nonumber\\
&+\half\sum_{k=1}^3\gamma_k\big\{(\Nbos_{\omega_k}+1)\mathcal{D}_{\hat{b}_k}[\rho] +\Nbos_{\omega_k}\mathcal{D}_{\hatd{b}_k}[\rho]\big\},
\end{align}
where the Lindblad superoperators are $\mathcal{D}_{\hat{o}}[\rho]\equiv 2\hat{o}\rho\hatd{o} -\hatd{o}\hat{o}\rho -\rho\hatd{o}\hat{o}$.
Here, $\omega_q$ is the TLS level splitting, while $\Gamma$ and $\widetilde\Gamma$ are the qubit relaxation and decoherence rate, respectively.
The mechanical damping rates are $\gamma_k\equiv \omega_k/Q$ with the quality factor $Q$.
The bosonic thermal occupation number at temperature $T$ is $\Nbos_\omega = (\exp\{\hbar\omega/k_{\rm B}T\}-1)^{-1}$ with the Boltzmann constant $k_{\rm B}$.
In our analysis, we assume that the mechanical modes as well as the TLS are initialized in their ground-state by some cooling mechanism~\cite{Abdi2017, Abdi2019} and a modulated electromagnetic drive excites the TLS.

\begin{figure}[tb]
\includegraphics[width=\columnwidth]{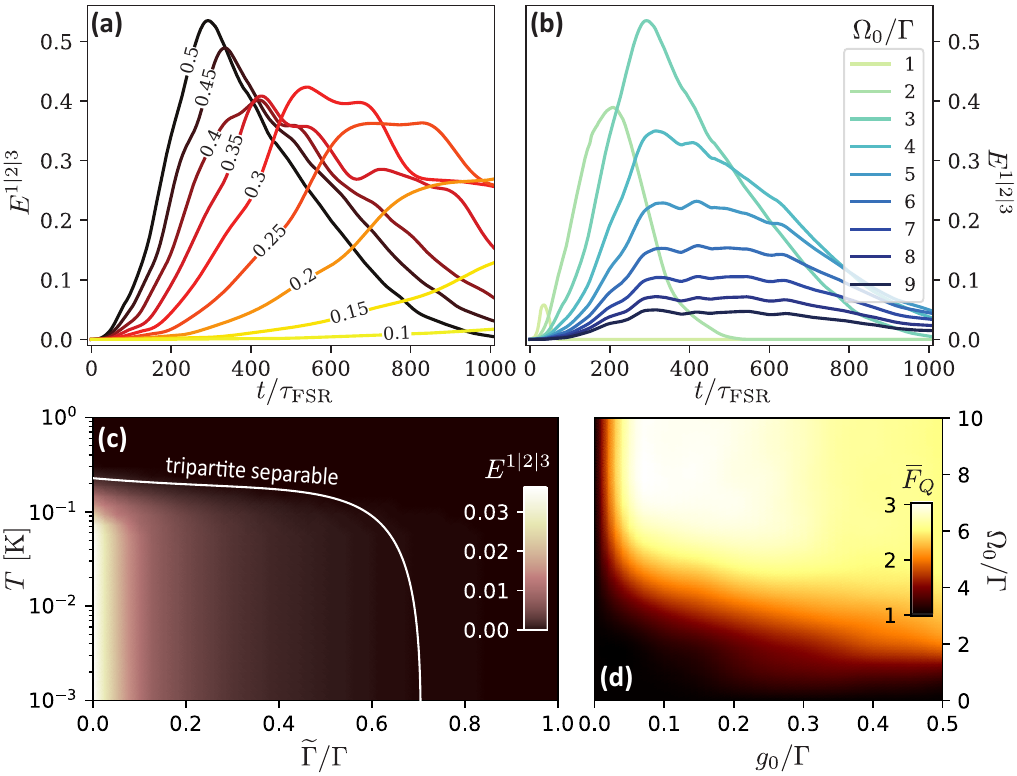}
\caption{%
Time evolution of the TPE (a) for different coupling rates (the curve labels show $g_0/\Gamma$) at the Rabi frequency $\Omega_0 =3\Gamma$, and (b) for different Rabi frequencies at $g_0=0.5\Gamma$.
(c) Robustness of longtime $E^{1|2|3}$: Density plot showing the tripartite entangled and separable parameter regions versus reservoir temperature and TLS pure dephasing rate $\widetilde\Gamma$ evaluated at $t=1000\tau_{\rm FSR}$ for $\Omega_0=7\Gamma$ and $g_0=0.5\Gamma$.
(d) The maximum normalized QFI as a function of coupling strength and drive amplitude.
}%
\label{fig:tri}%
\end{figure}

In numerical evaluation of Eq.~\eqref{master}, we consider a vibrational spectrum with $\delta_{\rm FSR}/2\pi=20$~MHz and $Q=10^7$ interacting with the TLS ($\omega_q/2\pi = 10$~GHz and $\Gamma/2\pi = 20$~MHz) with coupling rates $g_k =g_0 \lesssim \Gamma, \delta_{\rm FSR}$~\cite{Gokhale2020, Kervinen2020, Sletten2019, Arrangoiz2019}.
Unless specified otherwise, we take dilation refrigerator temperature of $T=10$~mK for the reservoir, set $\Delta = 5\Omega_0$ for getting $\mean{\sighat_z}\approx-1$ to ensure effective coupling between the mechanical modes, and assume an ideal TLS, $\widetilde\Gamma =0$.
By choosing an optimal value for the drive power $\Omega_0$ and coupling strengths, a long-lived tripartite mechanical entangled state is attained that sustains the qubit and thermal noise for thousands of the longest mechanical period, $\tau_{\rm FSR}\equiv 2\pi/\delta_{\rm FSR}$, which is hundreds of nanoseconds.

The results are summarized in Fig.~\ref{fig:tri}, where we present the genuine multipartite entanglement $E^{1|2|3}$ and the normalized quantum Fisher information $\overline{F}_{\!\!Q}$ as defined in Refs.~\cite{Adesso2008, Braunstein1994} and given in Appendix~\ref{sec:ngm}.
In Fig.~\ref{fig:tri}(a), time evolution of the entanglement in the three-mode system is plotted for different coupling rates, while in Fig.~\ref{fig:tri}(b) its evolution is shown for various drive amplitudes.
The entanglement curves exhibit a fast rise followed by a rapid decay for large coupling rates and/or low Rabi frequencies, which survives for several hundreds of oscillations.
Remarkably, in the opposite regime (weaker $g_0$ and/or higher $\Omega_0$) the system is dragged towards a quasi-stationary tripartite entanglement (TPE).
To study robustness of the entanglement against TLS pure dephasing as well as the thermal noise at higher temperatures, we take the $E^{1|2|3}$ for $g_0=0.5\Gamma$ and $\Omega_0=7\Gamma$ at $t=1000\tau_{\rm FSR}$ as a representative long-living TPE and show the effect of these two major sources of noise in Fig.~\ref{fig:tri}(c).
Interestingly, the system remains tripartite entangled, though fragile, for a wide range of noise parameters. In higher temperatures TLS thermalization is the main source of decoherence.
In a setup with a high enough $\omega_q$ TPE survives even up to $T \sim 10$~K, where the mechanical noises destroys the entanglement (not shown).
Finally, we compute the maximum QFI in measuring collective mechanical position $\hat{X}_3 = \sum_{i=1}^3\hat{x}_i$ in our scheme and study its dependence on the coupling strength $g_0$ and Rabi frequency $\Omega_0$ [Fig.~\ref{fig:tri}(d)].
The results suggest that the QFI increases by operating at high coupling strengths and drive amplitudes.
Interestingly, the maximum QFI even reaches the Heisenberg limit $\overline{F}_{\!\!Q}=3$, which means at least $1.73$ times enhancement in the sensitivity compared to a separable state, is obtained for moderate $g_0$ values.
It is worth mentioning here that the mechanical entangled state is non-Gaussian thanks to the nonlinear nature of the TLS. Note that non-Gaussianity is a crucial feature for universal quantum computation with continuous variable systems.
This is confirmed by computing the entropy distance of the density matrix from a reference Gaussian state as discussed in the Appendix~\ref{sec:ngm}.

%
%
%========================================%
\section{Multipartite entangled state}%
It is computationally expensive to study systems with larger number of modes through \eqref{master} since it becomes memory intensive.
Therefore, to expand our investigations to larger systems and testing our strategy for generating a multipartite entangled state, we employ an effective Gaussian model.
\begin{figure}[b]
\includegraphics[width=\columnwidth]{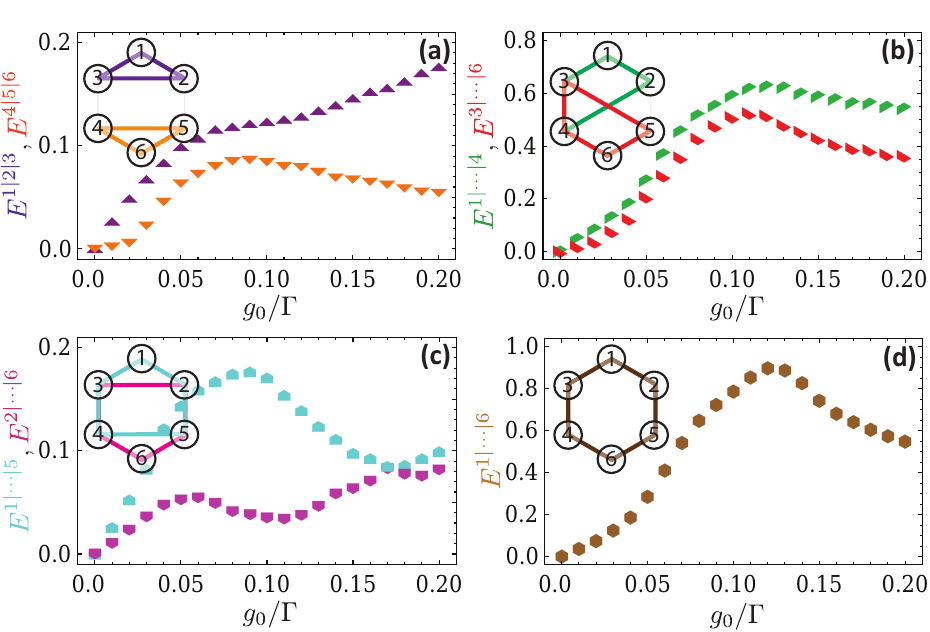}
\caption{%
Genuine $N$-partite entanglement for a set of six modes as a function of the coupling strength $g_0$ at $\Omega_0=3\Gamma$ and $t=1000\tau_{\rm FSR}$:
(a) $N=3$, (b) $N=4$, (c) $N=5$, and (d) $N=6$.
}
\label{fig:multi}%
\end{figure}
%
%We perform the numerical analysis on the multipartite entanglement in a set of $M$ vibrational modes in our proposed scheme.
The effective model is derived for both the Hamiltonian and the TLS-induced noise and damping on the mechanical modes through adiabatic elimination of the TLS.
Thanks to its Gaussian nature, the system is thus fully characterized by the covariance matrix (CM) $V$ when displaced to the mean values of the canonical operators.
To find the dynamics of CM, one forms a vector of quadrature operators $\mathbf{u}=[\hat{x}_1,\hat{p}_1,\cdots,\hat{x}_M,\hat{p}_M]^\intercal$ and derives the quantum Langevin equation of its elements, see Appendix~\ref{sec:mec} for the details.
In the compact form one gets $\dot{\bf u}=-A\mathbf{u}+\mathbf{n}$, where $\mathbf{n}$ is the vector of noise operators and the time-dependent drift matrix is given by 
\begin{equation}
A(t) = \hspace{-1mm}\left[\hspace{-1mm}
\begin{array}{ccccc}
	\half\kappa_1 & -\omega_1 & \cdots & 0 & 0 \\
	\omega_1+G_{1,1}(t) & \half\kappa_1 & \cdots & G_{1,M}(t) & 0 \\
	\vdots & \vdots & \ddots & \vdots & \vdots \\
	0 & 0 & \cdots & \half\kappa_M & -\omega_M \\
	G_{M,1}(t) & 0 & \cdots & \omega_M+G_{M,M}(t) & \half\kappa_M
\end{array}\hspace{-1mm}\right]\hspace{-1mm},
\label{diffus}
\end{equation}
where $G_{k,l}(t)\equiv \frac{1}{2\Delta(2\Nbos_{\omega_q}+1)}\frac{g_kg_l}{\omega_k\omega_l}\Omega(t)^2$ is the coupling strength of the modes to each other in the effective model.
Here, $\kappa_k = \gamma_k +\widetilde\gamma_k$ is the total damping rate composed of the intrinsic $\gamma_k$ and the TLS-induced damping rates $\widetilde\gamma_k \equiv g_k^2[\mathcal{S}(\omega_k) -\mathcal{S}(-\omega_k)]$ with $\mathcal{S}(\omega)$ the TLS steady-state fluctuation spectrum~\cite{Rabl2010a}.
%We remind that $A$ is time dependent through the modulated Rabi frequency.
The CM of the system is readily computed via the following equation
\begin{equation}
\dot{V}=AV+VA^\intercal +D,
\label{lyapunov}	
\end{equation}
where $D$ is the diffusion matrix, i.e. the matrix of noise correlators~\cite{Genes2008, Mari2009}.
In this work we assume that the noises are Gaussian and Markovian.
This brings us to $D=\bigoplus_{k=1}^MD_k$ with $D_k\equiv[\gamma_k(\Nbos_{\omega_k}+\half) +\widetilde\gamma_k(\tilde{n}_{\omega_k}+\half)]I_2$, where $I_2$ is a $2\times 2$ identity matrix.
The effective TLS-induced occupancy is calculated by $\tilde{n}_{\omega}\equiv \mathcal{S}(-\omega)/[\mathcal{S}(\omega) -\mathcal{S}(-\omega)]$.

The chosen set of parameters ensures validity of the TLS adiabatic elimination.
The convergence and reliability of the numerical results are verified as outlined in Appendix~\ref{sec:num}.
We restrict our analysis to $M=6$, as this is the maximum number of modes that we can afford given the computational resources.
The genuine $N$-partite entanglement values ($3\leq N\leq 6$) are evaluated at $t=1000\tau_{\rm FSR}$ as a function of $g_0$ and presented in Fig.~\ref{fig:multi}.
Two sets for $N=3,4,5$ cases are studied: The set of modes closest to the fundamental mode and those which gather around the farthest modes.
Since the effective coupling rate $G_{k,l}$ is proportional to the inverse mode frequencies, the fundamental mode has the strongest coupling to the active modes.
From this perspective, the stronger entanglement exhibited in the first set with an almost monotonic growth with the coupling rate is intuitive.
Therefore, the fundamental mode plays a central role in structure of the entangled cluster.
We should emphasize that the system is originally nonlinear, and thus, one expects non-Gaussian essence in the above discussed multimode mechanical states that is not perceptible in our Gaussian effective model.
\begin{figure}[tb]
\includegraphics[width=\columnwidth]{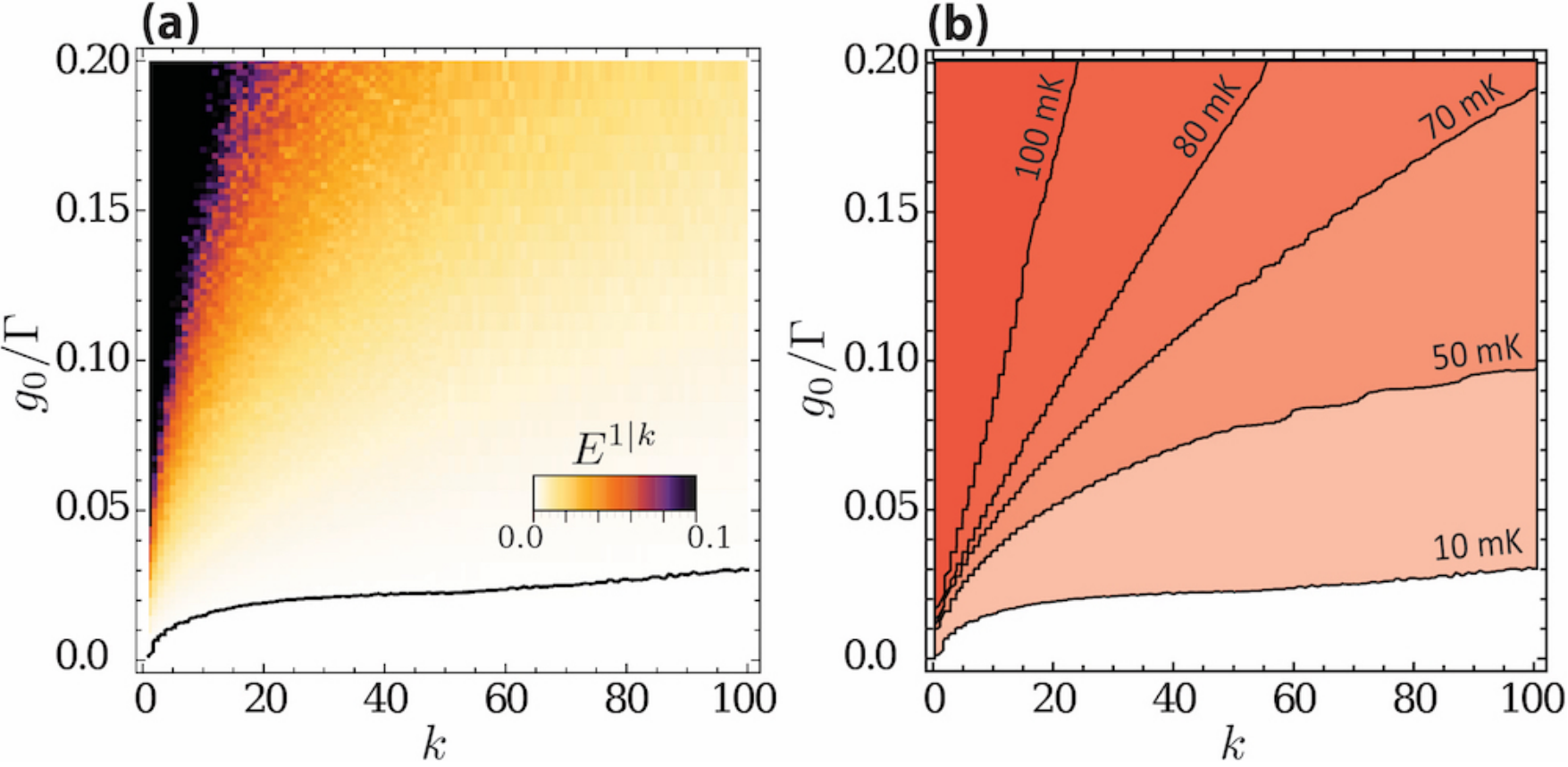}
\caption{%
The bipartite entanglement between the fundamental mode and the $k$th mode $E^{1|k}$ as a function of coupling rate $g_0$:
(a) Logarithmic negativity at $T=10$~mK.
(b) The entangled regions at five different temperatures.
In both plots the values correspond to $\Omega_0=3\Gamma$ at $t=1000\tau_{\rm FSR}$.
}
\label{fig:depth}%
\end{figure}
%

%
%
%========================================%
\section{Scalability}%
The fundamental mode is central in the entanglement structure of our scheme and at the same time it is the most vulnerable mode to the thermal noise.
Therefore, it is crucial to learn about the depth of its influence in the vibrational spectrum.
In other words, how strongly does it entangle to the higher order modes of the resonator and how robust it is.
%A property that we shall call \textit{entanglement depth}.
Hence, we single-out the fundamental and $k$th modes to study their bipartite entanglement $E^{1|k}$ [Fig.~\ref{fig:depth}(a)].
The logarithmic negativity~\cite{Plenio2005} of the bipartite system of $\rho_{\mathcal{H}_1\otimes\mathcal{H}_k}$ is evaluated for 100 modes at $t=1000\tau_{\rm FSR}$ as a function of $g_0$ when the qubit is excited by a drive modulated at $\omega_1$ and $\omega_k$.
In Fig.~\ref{fig:depth}(b) the residual entanglement, the parameter region where $E^{1|k}>0$, is presented at different temperatures.
The modes with lower frequencies assume stronger entanglement to the fundamental mode as expected.
Remarkably, by increasing the temperature they retain their entanglement quite firmly, with small change in the amount (not shown).
Meanwhile, the high-frequency modes preserve their entanglement with the fundamental mode, provided the coupling strength to the intermediating TLS $g_0$ is large enough.

%
%
%========================================%
\section{Summary and conclusion}%
To summarize, we have introduced and studied a compact and versatile device for creating a network of continuous variable multipartite entangled states.
The vibrational modes of a single resonator constitute nodes of the network.
These modes are coupled and entangled as a result of their interaction with a driven-dissipative two-level system.
We have proposed to activate a desired set of modes by modulating the TLS drive tone at their respective frequencies.
This results-in a long-living noise-resilient genuine multipartite entangled state that is useful for quantum metrology as the QFI of their state shows.
We have also studied the structure of entanglement for a hexagon network, proving the central role of the fundamental mode in the network.
Finally, influence of the fundamental mode on the mechanical spectrum shows that at the cryogenic temperatures hundreds of modes can join the network.

%
%
%----------ACKNOWLEDGEMENT----------%
\begin{acknowledgements}
%\textit{Acknowledgements.---}%
The author acknowledges Martin B. Plenio for careful reading and comments on the manuscript.
This work was supported by Iran National Science Foundation (INSF) via grant No.~98005028.
The support by STDPO and IUT through SBNHPCC is acknowledged.
\end{acknowledgements}

\appendix
%
%
%
%%%%%%%%%%%%%%%%%%%%%%%%%%%%%%%%%%
\section{Adiabatic elimination}\label{sec:adi}
\setcounter{equation}{0}
\setcounter{figure}{0}
\setcounter{table}{0}
\makeatletter
\renewcommand{\theequation}{A\arabic{equation}}
\renewcommand{\thefigure}{A\arabic{figure}}
In this appendix we present the details of our procedure in deriving the effective Hamiltonian \eqref{eff} starting from the original Hamiltonian \eqref{ham}.
We start by removing the qubit-mechanical interaction by applying the polaron transformation $\hat{U}=\exp\{\half i\sighat_z\hat{P} \}$ with the collective dimensionless momentum~\cite{Rabl2011}
\begin{equation}
\hat{P}=-i\sum_k\frac{g_k}{\omega_k}(\hat{b}_k-\hatd{b}_k).
\end{equation}
We are then brought to
\begin{equation}
\hat{U}\hat{H}\hatd{U}=\frac{\Delta}{2}\sighat_z+\frac{\Omega}{2}\big(\sighat_+e^{-i\hat{P}}+\sighat_- e^{+i\hat{P}}\big)
+\sum_k\omega_k\hatd{b}_k\hat{b}_k,
\label{polaron}
\end{equation}
where we have discarded a constant term.

The adiabatic elimination is performed by assuming that the qubit decay rate is larger than its coupling strength to any of the mechanical modes $\Gamma \gg \{g_k\}$.
In this regime the total density matrix can be approximated by $\rho(t)\simeq\rho_{\rm TLS}(t)\otimes \rho_{\rm Mec}(t)$.
Therefore, a mean-field approximation becomes applicable.
Then the following equations describe the qubit dynamics~\cite{Carmichael1999}
\begin{subequations}
\begin{align}
\!\mean{\dot{\sighat}_+} &=-\Big(\frac{\Gamma}{2}(2\Nbos_{\omega_q}+1) -i\Delta\Big) \mean{\sighat_+} -\half i\Omega e^{+i\mean{\hat{P}}}\mean{\sighat_z}, \\
\!\mean{\dot{\sighat}_-} &=-\Big(\frac{\Gamma}{2}(2\Nbos_{\omega_q}+1) +i\Delta\Big) \mean{\sighat_-} +\half i\Omega e^{-i\mean{\hat{P}}}\mean{\sighat_z}, \\
\!\mean{\dot{\sighat}_z} &=-\Gamma\Big(\one +(2\Nbos_{\omega_q}+1)\mean{\sighat_z}\Big) \nonumber\\
&~~~~~-i\Omega\big(\mean{\sighat_+} e^{-i\mean{\hat{P}}} -\mean{\sighat_-} e^{+i\mean{\hat{P}}}\big),
\end{align}%
\label{qubit}%
\end{subequations}%
where for the sake of simplicity an ideal qubit with no pure dephasing $\widetilde\Gamma = 0$ is assume.
The solutions to the above equations in qubits steady-state, i.e. the states with $\mean{\sighat_\pm}_{\rm ss}=\mean{\sighat_z}_{\rm ss}=0$, are:
\begin{subequations}
\begin{align}
\mean{\sighat_\pm}_{\rm ss} &= \frac{\pm i\Omega\Big(\Gamma(2\Nbos_{\omega_q}+1) \mp2i\Delta\Big)e^{\pm i\mean{\hat{P}}}}{(2\Nbos_{\omega_q}+1)\Big(\Gamma^2(2\Nbos_{\omega_q}+1)^2 +4\Delta^2 +2\Omega^2\Big)}, \\
\mean{\sighat_z}_{\rm ss} &= \frac{-\Gamma^2(2\Nbos_{\omega_q}+1)^2 +4\Delta^2}{(2\Nbos_{\omega_q}+1)\Big(\Gamma^2(2\Nbos_{\omega_q}+1)^2 +4\Delta^2 +2\Omega^2\Big)}.
\end{align}
\label{qss}
\end{subequations}
A simple set of equations is found by operating the system at the far detuned regime where $\Delta \gg \Gamma, \Omega$
\begin{subequations}
\begin{align}
\mean{\sighat_\pm}_{\rm ss} &\approx \frac{\Omega}{2\Delta(2\Nbos_{\omega_q}+1)}e^{\pm i\mean{\hat{P}}}, \\
\mean{\sighat_z}_{\rm ss} &\approx -1.
\end{align}
\end{subequations}
By plugging these in \eqref{polaron} for the qubit operators we arrive at
\begin{equation}
\hat{H}_{\rm ad} = \sum_k\omega_k \hatd{b}_k\hat{b}_k 
+\frac{\Omega^2}{2\Delta(2\Nbos_{\omega_q}+1)}\cos(\hat{P}-\mean{\hat{P}}),
\label{adiabatic}
\end{equation}
after discarding a constant shift in the energy.
This Hamiltonian is valid for the above discussed regime and adequately addresses the system dynamics after the TLS passes its transient dynamics $t>\Gamma^{-1}$.
The collective momentum then is expected to exhibit small fluctuations around its mean value $\tr{}{(\hat{P}-\mean{\hat{P}})\rho_{\rm Mec}(t)} \ll 1$.
This allows us to expand the cosine and keep up to the quadratic terms
\[
\cos(\hat{P}-\mean{\hat{P}})\approx 1-\half(\hat{P}-\mean{\hat{P}})^2 = 1 -\half\big(\hat{P}^2+\mean{\hat{P}}^2 -2\mean{\hat{P}}\hat{P}\big).
\]
The second term in the parentheses gives a constant energy, while the third causes a shift in the phase space. We skip both and write
\begin{equation}
\hat{H}_{\rm ad} \approx \sum_k \omega_k\hatd{b}_k\hat{b}_k +\frac{1}{2}\sum_{k,l}G_{k,l}\Big(\hat{b}_k\hat{b}_l-\hat{b}^{}_k\hatd{b}_l +\text{H.c.}\Big),
\label{adapp}
\end{equation}
where the effective coupling rate is
\begin{equation}
G_{k,l}\equiv \frac{\Omega^2}{2\Delta(2\Nbos_{\omega_q}+1)}\frac{g_k g_l}{\omega_k\omega_l}.
\label{coupling}
\end{equation}

\begin{figure}[b]
\includegraphics[width=0.7\columnwidth]{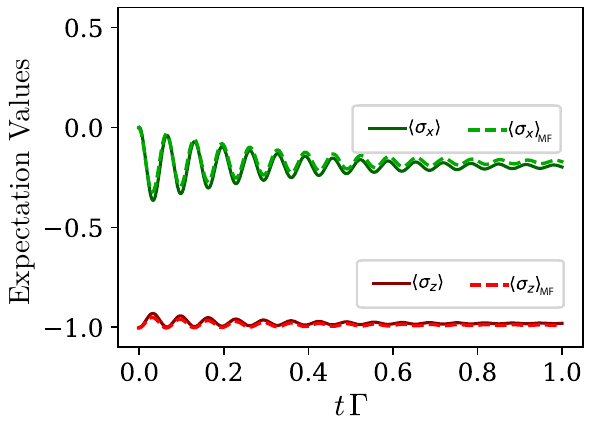}
\caption{%
Comparison of the expectations values for the TLS operators: the exact (solid lines) versus mean-field approximation (dashed lines), which are solutions to Eqs.~\eqref{qubit}. See the text for details.}
\label{fig:qubit}
\end{figure}

To check the validity of mean-field approximation in the discussed regime, we numerically solve the qubit equations in \eqref{qubit} replacing $\hat{P}$ with a real number and compare the results with the solution of quantum optical master equation \eqref{master}.
In Fig.~\ref{fig:qubit} the two solutions are compared against each other. One clearly verifies that the solutions are very close in the adiabatic elimination regime.

On the dissipation side of the dynamics, the TLS-mechanical interaction causes extra dissipation in the mechanical modes.
This is formulated effectively through second order perturbation theory~\cite{Jaehne2008}.
The additional damping rate for each mechanical mode is then $\widetilde\gamma_k \equiv g_k^2[\mathcal{S}(\omega_k) -\mathcal{S}(-\omega_k)]$.
According to the quantum dissipation-fluctuation theorem there is an accompanying thermal noise with occupation number $\tilde{n}_{\omega_k}\equiv \mathcal{S}(-\omega_k)/[\mathcal{S}(\omega_k) -\mathcal{S}(-\omega_k)]$.
Here, the TLS steady-state fluctuation spectrum is given by
\begin{equation}
\mathcal{S}(\omega)=\half\Re\int_0^\infty\hspace{-2mm} ds \big[\mean{\sighat_z(s)\sighat_z(0)}_{\rm ss}-\mean{\sighat_z}_{\rm ss}^2\big]e^{i\omega s}.
\end{equation}
The spectrum is evaluated using quantum regression theorem~\cite{Carmichael1999}.
The resulting expression is cumbersome. Therefore, we only summarize the equations that lead to it in the same lines of Ref.~\cite{Rabl2010a}.
We express the spectrum of TLS as $\mathcal{S}(\omega)=\half\Re\{C_3(s=-i\omega)\}$, where $C_3$ is the third element of the following vector
\begin{equation}
\mathbf{C}(s) = (sI -M)^{-1}\mathbf{v},
\end{equation}
where $I$ is a $3\times 3$ identity matrix and the matrix $M$ and vector $\mathbf{v}$ are defined through Eqs.~\eqref{qubit} and their steady-state solutions as
\begin{widetext}
\begin{equation}
M = \left(
\begin{array}{ccc}
	-\half\Gamma(2\Nbos_{\omega_q}+1) +\widetilde\Gamma +i\Delta & 0 & -\half i\Omega \\
	0 & -\half\Gamma(2\Nbos_{\omega_q}+1) +\widetilde\Gamma -i\Delta & +\half i\Omega \\
	-i\Omega & i\Omega & -\Gamma(2\Nbos_{\omega_q}+1) \\
\end{array}\right),\hspace{3mm}
\mathbf{v}=\left(\begin{array}{c}
	-\mean{\sighat_+}_{\rm ss}(1+\mean{\sighat_z}_{\rm ss}) \\
	+\mean{\sighat_-}_{\rm ss}(1-\mean{\sighat_z}_{\rm ss}) \\
	1-\mean{\sighat_z}_{\rm ss}^2 \\
\end{array}\right).
\end{equation}
\end{widetext}
These TLS-induced damping and noises are taken into account in studying the dynamics of effective fully mechanical system, see Appendix~\ref{sec:mec}.

%%%%%%%%%%%%%%%%%%%%%%%%%%%%%%%%%%
\section{Rotating Wave Approximation}\label{sec:rwa}
\setcounter{equation}{0}
\setcounter{figure}{0}
\setcounter{table}{0}
\makeatletter
\renewcommand{\theequation}{B\arabic{equation}}
\renewcommand{\thefigure}{B\arabic{figure}}
The Hamiltonian \eqref{adapp} provides us with interactions of two-mode squeezing and state-transfer between all modes, and thus, is potentially rich for quantum information processing purposes. However, none of the above interactions are on resonance.
This becomes clear by moving to the interaction picture of $\hat{H}_0 = \sum_k\omega_k\hatd{b}_k\hat{b}_k$.
In this appendix we show that a selected set of modes can resonantly brought into desired interaction by applying a modulated drive to the qubit.
In other words, by replacing the $\Omega$ with
\begin{equation}
\Omega(t)=\Omega_0\sum_i\cos w_i t,
\label{drive}
\end{equation}
with properly chosen modulation frequencies $w_i$ that will become clear, shortly.
To show this, we move to the interaction picture of $\hat{H}_0$ and arrive at
\[
\widetilde{H}_{\rm ad} \approx \frac{1}{2}\sum_{k,l}G_{k,l}(t)\big[\hat{b}_k\hat{b}_l e^{-i(\omega_k+\omega_l)t} -\hat{b}_k^{}\hatd{b}_l e^{-i(\omega_k-\omega_l)t} +\text{H.c.}\big],
\]
where now $G_{k,l}$ are time dependent through the Rabi frequency, see Eq.~\eqref{coupling}.
The drive in \eqref{drive} then gives us
\begin{align*}
\Omega(t)^2 &= \Omega_0^2 \sum_{i,j}\cos (w_it)\cos (w_jt) \\
&=\half\Omega_0^2\sum_{i,j}\Big\{\cos\!\big[(w_i+w_j)t\big] +\cos\!\big[(w_i-w_j)t\big] \Big\}\\
&=\tfrac{1}{4}\Omega_0^2\sum_{i,j}\Big[e^{i(w_i+w_j)t} +e^{i(w_i-w_j)t} +\text{c.c.}\Big].
\end{align*}
In the numerical analysis that is performed in the manuscript for multimode systems larger than three we use Hamiltonian \eqref{adapp} and derive the Langevin equations, see Appendix~\ref{sec:mec}.
Nonetheless, it is instructive to see that how a time dependent drive can link the mechanical modes to each other.
Therefore, by setting the above equation back in $\widetilde{H}_{\rm ad}$ and performing a rotating wave approximation (RWA), only few of the mode interactions can survive.
When the mode frequencies are \textit{non-commensurate} the desired set of modes with frequencies $\{\omega_k\}$ can be \textit{activated} by simply setting $w_k=\omega_k$ for any set of modes.
This gives
\begin{equation}
\hat{H}_{\rm RWA}=\frac{1}{2}\sum_{k, l}^M G_{k,l}\hat{p}_k\hat{p}_l,
\label{rwanon}
\end{equation}
where $G_{k,l}$ is the same as \eqref{coupling} but with $\Omega$ replaced by $\Omega_0$.
\begin{figure}[b]
\includegraphics[width=\columnwidth]{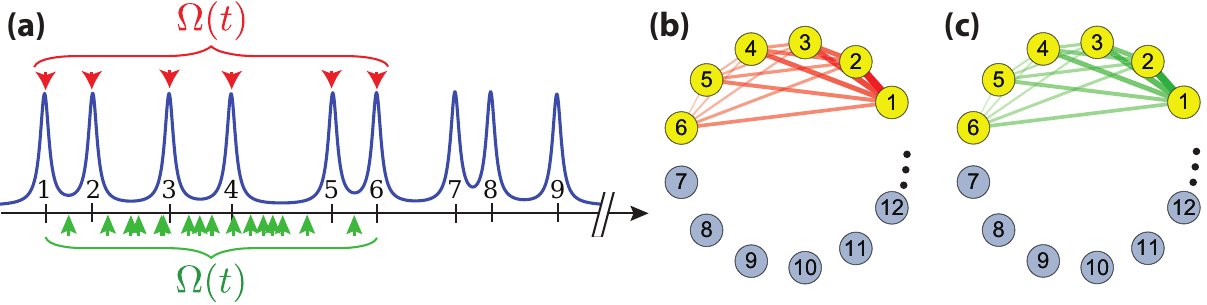}
\caption{%
(a) When the spectrum is non-Commensurate any desired set of modes can be activated by simply modulating the TLS drive on their frequencies (red arrows) or at half of the sum frequencies (green arrows).
The activated modes form a complete graph, though weighted, for the two-mode squeezing interaction in both modulation strategies as illustrated for in: (b) mode frequencies modulation and (c) half of the sum frequencies modulation.}
\label{fig:nonC}
\end{figure}
Note that the interaction with and among the rest of spectrum become counter-rotating and thus are ignored.
An alternative modulation strategy is to drive at half of the sum frequencies such that $w_i=\omega_{k,l}\coloneqq \half(\omega_k +\omega_l)$. In this case the effective Hamiltonian in RWA is in the following form
\begin{equation}
\hat{H}_{\rm RWA}'=\frac{1}{4}\sum_{k, l}^M G_{k,l}(\hat{x}_k\hat{x}_l +\hat{p}_k\hat{p}_l),
\end{equation}
Concerning the two-mode squeezing interaction, both schemes result the same interaction graphs as it is conceived from Fig.~\ref{fig:nonC}.

\begin{figure*}[t]
\includegraphics[width=0.8\textwidth]{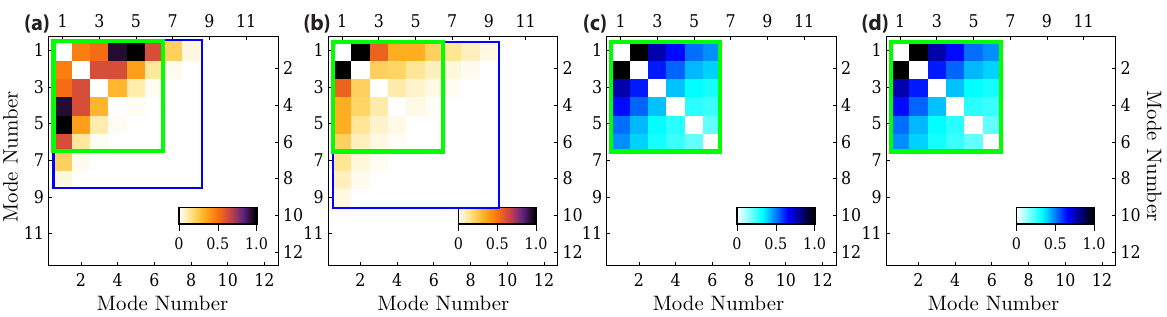}
\caption{%
Normalized matrix elements of the effective two-mode squeezing coupling matrices $\mathcal{F}_{k,l}$:
(a) and (b) for commensurate spectrum with modulations at half sum frequencies and the mode frequencies, respectively.
In (c) and (d) the same are shown for a noncommensurate spectrum.
The green square highlights the \textit{target} set of modes, whereas the blue square illustrates the set of \textit{activated} modes that are directly coupled to the target set.}
\label{fig:matrix}
\end{figure*}

Nevertheless, in our proposed setup the set of mode frequencies $\{\omega_k\}$ are commensurate. This indeed involves a few undesired modes in the system dynamics.
%Again, by choosing $w_k=\omega_k$ for the sequence of the $M$ lowest frequency modes, all modes up to the ($2M-1$)th mode get activated.
%The number of `undesired modes' can be reduced by smarter choices for $w_k$. However, for the sake of simplicity we stick the case of $w_k=\omega_k$.
Another cause of the commensurate spectrum is that the coupling rate coefficients become weighted as some interactions receive drive from other drive tones.
%Remarkably, the state-transfer network spreads to a wider set of modes.
Therefore, the RWA Hamiltonian in our setup becomes
\begin{equation}
\hat{H}_{\rm RWA} =\half\sum_{k,l}G_{k,l}\big(B^{\rm tms}_{k,l}\hat{b}_k\hat{b}_l -B^{\rm qst}_{k,l}\hat{b}_k\hatd{b}_l +\text{H.c.}\big),
\label{rwa}
\end{equation}
where $B^{\rm tms}$ and $B^{\rm qst}$ are the adjacency matrices of the two-mode squeezing and quantum state-transfer interactions, respectively.
%The Hamiltonian \eqref{rwa} is rewritten in terms of Hermitian position and momentum operators as
%\begin{equation}
%\hat{H}_{\rm RWA} 
%= \half\sum_{k,l} \big(G_{k,l}^x\hat{x}_k\hat{x}_l -G_{k,l}^p\hat{p}_k\hat{p}_l\big),
%\label{rwa2}
%\end{equation}
%with the coupling rates
%\begin{align*}
%G_{k,l}^x &\equiv \mathcal{F}_{k,l}-\mathcal{F}'_{k,l}, \\
%G_{k,l}^p &\equiv \mathcal{F}_{k,l}+\mathcal{F}'_{k,l}.
%\end{align*}
%In Fig.~\ref{fig:graphs} the interaction matrices $G^x$ and $G^p$ illustrated.
Fig.~\ref{fig:matrix} shows the weight of each interaction via an illustrative presentation of the matrix elements of the adjacency matrix $\mathcal{F}_{k,l}\equiv G_{k,l}B^{\rm tms}_{k,l}$ for two different modulation schemes described above.
We notice that even though the modulated drive is tuned for a set of target modes (the green square), the other modes are also connected (the blue square).
By taking into account the inverse frequencies that appear in the interaction strengths, the effect of these higher modes becomes negligible.
This indeed is also shown as the graphs in Fig.~\ref{fig:scheme} in the main text.

%%%%%%%%%%%%%%%%%%%%%%%%%%%%%%%%%%
\section{Mechanical Langevin equations}\label{sec:mec}
\setcounter{equation}{0}
\setcounter{figure}{0}
\setcounter{table}{0}
\makeatletter
\renewcommand{\theequation}{C\arabic{equation}}
\renewcommand{\thefigure}{C\arabic{figure}}
The Hamiltonian \eqref{adapp} is Gaussian and allows us to study the system dynamics through the covariance matrix of the mechanical canonical operators.
The quantum Langevin equations (QLEs) include the effect of environment on dynamics of the system operators are used to obtain time evolution of the covariance matrix.
We find the following QLEs for the motion of the mechanical modes:
\begin{equation}
\dot{b}_k =-(\half\kappa_k +i\omega_k)\hat{b}_k-i\sum_l G_{k,l}(t)(\hat{b}_l +\hatd{b}_l) +\hat{\xi}_k,
\label{langevin}
\end{equation}
where $\kappa_k \equiv \gamma_k+\widetilde\gamma_k$ is the total damping rate stemming from the support, $\gamma_k$, and the TLS decoherences, $\widetilde\gamma_k$, as discussed in Appendix~\ref{sec:adi}.
The total noise operator is also divided into the support noise and the TLS induced noise $\hat{\xi}_k \equiv \sqrt{\gamma_k}\hat{b}_k^{\rm in} +\sqrt{\widetilde\gamma_k}\tilde{b}_k^{\rm in}$.
%Remarkably, in the working regime that has been studied in this work the pure dephasing performs as a Gaussian noise.
Assuming that the intrinsic motional noise is Markovian, the noise operator has the following non-vanishing correlation functions
\begin{subequations}
\begin{align}
\hspace{-2.7mm}\mean{\hat\xi^{}_k(t)\hatd\xi_l(t')} &= \big[\gamma_k(\Nbos_{\omega_k}+1) +\widetilde\gamma_k(\widetilde{n}_{\omega_k}+1)\big]\delta(t-t')\delta_{kl}, \\
\hspace{-2.7mm}\mean{\hatd\xi_k(t)\hat\xi^{}_l(t')} &= \big[\gamma_k\Nbos_{\omega_k} +\widetilde\gamma_k\widetilde{n}_{\omega_k}\big]\delta(t-t')\delta_{kl},
\end{align}
\end{subequations}
By forming a vector of Hermitian operators $\mathbf{u}\equiv [\hat{x}_1,\hat{p}_1,\cdots,\hat{x}_M,\hat{p}_M]^\intercal$, one arrives at the following compact form of Eqs.~\eqref{langevin}
\begin{equation}
\dot{\bf u} = A\mathbf{u} +\mathbf{n},
\label{langcom}
\end{equation}
where $A$ is given by \eqref{diffus} and $\mathbf{n}=[\hat\xi^x_1,\hat\xi^p_1,\cdots,\hat\xi^x_M,\hat\xi^p_M]$ is the vector of noise operators with $\hat\xi_k\equiv (\hat\xi_k^x+i\hat\xi_k^p)/\sqrt{2}$.
The covariance matrix dynamics is readily calculated from \eqref{langcom} as Eq.~\eqref{lyapunov} in the main text~\cite{Mari2009}.

%%%%%%%%%%%%%%%%%%%%%%%%%%%%%%%%%%
\section{Numerical method}\label{sec:num}
\setcounter{equation}{0}
\setcounter{figure}{0}
\setcounter{table}{0}
\makeatletter
\renewcommand{\theequation}{D\arabic{equation}}
\renewcommand{\thefigure}{D\arabic{figure}}
For performing the numerical integration over the quantum optical master equation \eqref{master} we use the QuTiP package~\cite{Johansson2012}.
The Hilbert space $\mathcal{H}$ truncation was done by minimizing the error.
Our estimation for the error by tracking the convergence curve shows that for a truncation at $\text{dim}\{\mathcal{H}\}=2\times10\times9\times8$ an approximate error of $\approx 1.73\%$ is committed, which is the minimum possible error given the computational resources at hand.

\begin{figure*}[tb]
\includegraphics[width=1.4\columnwidth]{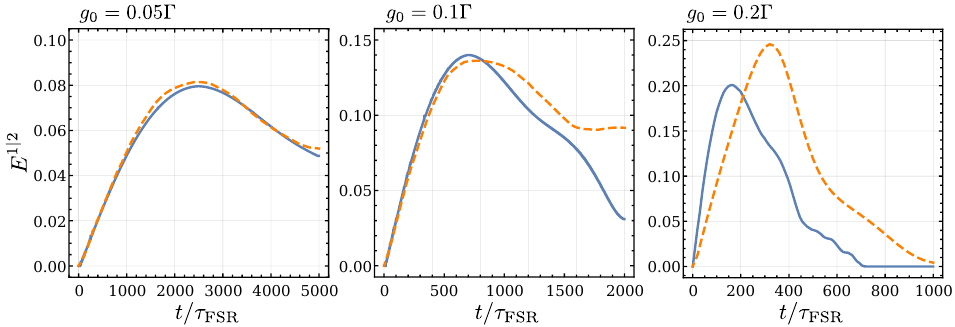}
\caption{%
Comparison of the entanglement $E^{1|2}$ as an illustration of the consistency: the original Hamiltonian (dashed orange) versus the effective Hamiltonian (solid blue) for three different coupling strengths.
Here, $\Omega_0=1\Gamma$ and the other parameters are the same as in the main text.}
\label{fig:compare}
\end{figure*}

The study of the original Hamiltonian with more than three mechanical modes is not feasible because of the huge demands for the computational memories.
We, therefore, have turned to use the Gaussian effective Hamiltonian which is valid in the regime discussed in Appendices~\ref{sec:adi} and \ref{sec:mec}.
To numerically prove its validity, we compare the results obtained from \eqref{master} and \eqref{lyapunov}.
In Fig.~\ref{fig:compare} the bipartite entanglement $E^{1|2}$ obtained from the original and effective models are plotted for three different coupling strength values.
The results confirms that the behavior of both models coincides for weak coupling strengths.
This is indeed as one would expect as the adiabatic elimination is valid when the TLS arrives at its quasi-steady state.
The deviation emerges as the coupling rate assumes higher values.
Such that the effective model overestimates the TLS-induced decoherence.

%%%%%%%%%%%%%%%%%%%%%%%%%%%%%%%%%%
\section{Entanglement, quantum Fisher information, and non-Gaussianity}\label{sec:ngm}
\setcounter{equation}{0}
\setcounter{figure}{0}
\setcounter{table}{0}
\makeatletter
\renewcommand{\theequation}{E\arabic{equation}}
\renewcommand{\thefigure}{E\arabic{figure}}
In this appendix we deal with the details of the entanglement properties, represent the measures used in the main text, and study the non-Gaussianity of the entangled state.

In order to analyze the entanglement structure of the system, we first show that our scheme indeed leads to a GHZ-state.
For the sake of clarity and simplicity the case of noncommensurate modes for which the interaction is pure $\hat{p}_k\hat{p}_l$ is discussed, see Eq.~\eqref{rwanon}.
The results are easily generalized to a system with commensurate mode spectrum.
In the interaction picture, after a time interval $t$ the Hamiltonian \eqref{rwanon} transforms the canonical operators as
\begin{subequations}
\begin{align}
&\hat{x}_k\rightarrow\hat{x}_k-t\sum_l G_{k,l}\hat{p}_l, \\
&\hat{p}_k\rightarrow\hat{p}_k.
\end{align}
\end{subequations}
The effective coupling constant $G_{k,l}$ is inversely proportional to the mode frequencies $G_{k,l}\propto (\omega_k\omega_l)^{-1}$.
Therefore, the mode with the lowest frequency couples more strongly to the other nodes in the network [Fig.~\ref{fig:nonC}].
This suggests a star-shaped graph for the network with the lowest mode at the center.
Therefore, the position transformations can be approximately simplified to 
$$\hat{x}_1\rightarrow \hat{x}_1-\sum_k \chi_k\hat{p}_k$$
for the `first' mode and
$$\hat{x}_k\rightarrow\hat{x}_k-\chi_k\hat{p}_1$$
for the rest with $\chi_k\equiv tG_{1,k}$.
The entanglement of a state is not affected by local unitary transformations.
Hence, after a local $\pi/2-$rotation on the first-mode one gets $\hat{x}_1\to\hat{p}_1$ and $\hat{p}_1\to-\hat{x}_1$.
Then it is easy to verify that state of the system at a given time $t$ becomes such that
\begin{subequations}
\begin{align}
&\sum_{k=1}^M\hat{p}_k(t) = \sum_{k=1}^M\hat{p}_k(0)e^{-\chi_k}, \\
&\hat{x}_k(t)-\hat{x}_l(t)=\hat{x}_k(0)e^{-\chi_k}-\hat{x}_l(0)e^{-\chi_l},
\end{align}
\end{subequations}
with $(k,l=1,2,\cdots M)$.
After a long enough drive one has $\chi_k\rightarrow\infty$, where the system reaches a Greenberger-Horne-Zeilinger entangled state~\cite{Pfister2004, Zhang2006}.

We use the genuine multipartite entanglement measure introduced in Ref.~\cite{Adesso2008} to quantify the entanglement in our system.
The measure exploits monogamy property of the entanglement such that
$$E^{1|2,3,\cdots,M} =\sum_{j=2}^M E^{1|j} +\sum_{k>j}^M\sum_{j=2}^M E^{1|j|k} +\cdots +E^{\underline{1}|2|\cdots|M},$$
where the underline denotes the \textit{focus} party and $E$ is a proper measure of entanglement~\cite{Adesso2014}.
The genuine residual $N$-partite entanglement is then calculated as the minimum over all permutations of the subsystem indices
\begin{equation}
E^{1|2|\cdots|M} \equiv \min\{E^{\underline{i_1}|i_2|\cdots|i_M}\}.
\end{equation}

The quantum Fisher information determines the Cramer-Rao bound in parameter estimation and saturates to the Heisenberg limit for a fully entangled system~\cite{Giovannetti2011}.
Hence, it reflects the degree of the multipartite entanglement~\cite{Krischek2011, Hyllus2012, Toth2012}.
For a mixed state $\rho$ and observable $\hat{O}$ the QFI is defined as
\begin{equation}
F_Q[\rho,\hat{O}]=2\sum_{k,l}|\bra{k}\hat{O}\ket{l}|^2\frac{(\lambda_k-\lambda_l)^2}{\lambda_k+\lambda_l},
\end{equation}
where $\lambda_k$ and $\ket{k}$ are the eigenvalues and eigenvectors of $\rho$, respectively.
The sum is over indices that $\lambda_k+\lambda_l>0$~\cite{Braunstein1994}.
Here, we take the collective mechanical position $\hat{X}_M\equiv \sum_{k=1}^M\hat{x}_k$ as the observable and introduce $\overline{F}_{\!Q}\equiv \frac{1}{M}F_Q$, the normalized QFI.
This quantity is then upperbounded by $M$ for a fully entangled system.

\begin{figure}[b]
\includegraphics[width=0.8\columnwidth]{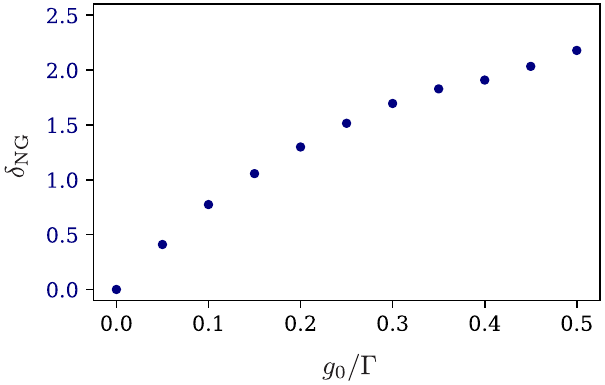}
\caption{%
The non-Gaussianity of the triangle system versus coupling strength.
The parameters are given in the main text. Here, we set $\Omega_0 = 3\Gamma$ and evaluate the measure of non-Gaussianity at $t=1000\tau_{\rm FSR}$.}
\label{fig:ng3}
\end{figure}

Finally, in order to examine the non-Gaussian nature of the entangled state in the triangle system we compute the non-Gaussianity measure of the state using the one introduced in Ref.~\cite{Genoni2010}: $\delta_{\rm NG}$ of a given state $\rho$ is defined as the distance of its entropy from a reference Gaussian state $\rho_{\rm G}$: $\delta_{\rm NG}\equiv S(\rho_{\rm G})-S(\rho)$ where $S(\rho)=-\text{Tr}\{\rho\log\rho\}$ is the von Neumann entropy.
The measure of non-Gaussianity is computed for the mechanical state $\rho_{\rm m}\equiv\text{Tr}_{\textsc{tls}}\{\rho\}$, where $\text{Tr}_{\textsc{tls}}$ denotes partial trace over the TLS.
In Fig.~\ref{fig:ng3} the measure is shown as a function of coupling strength $g_0$ for the triangle system. As expected, the non-Gaussianity of state increases with the coupling rate.

%
%
%----------REFERENCES----------%
\bibliography{CVGHZ.bib}

\begin{thebibliography}{95}%
\makeatletter
\providecommand \@ifxundefined [1]{%
 \@ifx{#1\undefined}
}%
\providecommand \@ifnum [1]{%
 \ifnum #1\expandafter \@firstoftwo
 \else \expandafter \@secondoftwo
 \fi
}%
\providecommand \@ifx [1]{%
 \ifx #1\expandafter \@firstoftwo
 \else \expandafter \@secondoftwo
 \fi
}%
\providecommand \natexlab [1]{#1}%
\providecommand \enquote  [1]{``#1''}%
\providecommand \bibnamefont  [1]{#1}%
\providecommand \bibfnamefont [1]{#1}%
\providecommand \citenamefont [1]{#1}%
\providecommand \href@noop [0]{\@secondoftwo}%
\providecommand \href [0]{\begingroup \@sanitize@url \@href}%
\providecommand \@href[1]{\@@startlink{#1}\@@href}%
\providecommand \@@href[1]{\endgroup#1\@@endlink}%
\providecommand \@sanitize@url [0]{\catcode `\\12\catcode `\$12\catcode
  `\&12\catcode `\#12\catcode `\^12\catcode `\_12\catcode `\%12\relax}%
\providecommand \@@startlink[1]{}%
\providecommand \@@endlink[0]{}%
\providecommand \url  [0]{\begingroup\@sanitize@url \@url }%
\providecommand \@url [1]{\endgroup\@href {#1}{\urlprefix }}%
\providecommand \urlprefix  [0]{URL }%
\providecommand \Eprint [0]{\href }%
\providecommand \doibase [0]{https://doi.org/}%
\providecommand \selectlanguage [0]{\@gobble}%
\providecommand \bibinfo  [0]{\@secondoftwo}%
\providecommand \bibfield  [0]{\@secondoftwo}%
\providecommand \translation [1]{[#1]}%
\providecommand \BibitemOpen [0]{}%
\providecommand \bibitemStop [0]{}%
\providecommand \bibitemNoStop [0]{.\EOS\space}%
\providecommand \EOS [0]{\spacefactor3000\relax}%
\providecommand \BibitemShut  [1]{\csname bibitem#1\endcsname}%
\let\auto@bib@innerbib\@empty
%</preamble>
\bibitem [{\citenamefont {Bennett}\ \emph {et~al.}(1993)\citenamefont
  {Bennett}, \citenamefont {Brassard}, \citenamefont {Cr{\'e}peau},
  \citenamefont {Jozsa}, \citenamefont {Peres},\ and\ \citenamefont
  {Wootters}}]{Bennett1993}%
  \BibitemOpen
  \bibfield  {author} {\bibinfo {author} {\bibfnamefont {C.~H.}\ \bibnamefont
  {Bennett}}, \bibinfo {author} {\bibfnamefont {G.}~\bibnamefont {Brassard}},
  \bibinfo {author} {\bibfnamefont {C.}~\bibnamefont {Cr{\'e}peau}}, \bibinfo
  {author} {\bibfnamefont {R.}~\bibnamefont {Jozsa}}, \bibinfo {author}
  {\bibfnamefont {A.}~\bibnamefont {Peres}},\ and\ \bibinfo {author}
  {\bibfnamefont {W.~K.}\ \bibnamefont {Wootters}},\ }\bibfield  {title}
  {\bibinfo {title} {{``Teleporting an Unknown Quantum State via Dual Classical
  and Einstein--Podolsky--Rosen Channels''}},\ }\href
  {https://doi.org/10.1103/PhysRevLett.70.1895} {\bibfield  {journal} {\bibinfo
   {journal} {Phys. Rev. Lett.}\ }\textbf {\bibinfo {volume} {70}},\ \bibinfo
  {pages} {1895} (\bibinfo {year} {1993})}\BibitemShut {NoStop}%
\bibitem [{\citenamefont {Braunstein}\ and\ \citenamefont
  {Kimble}(1998)}]{Braunstein1998}%
  \BibitemOpen
  \bibfield  {author} {\bibinfo {author} {\bibfnamefont {S.~L.}\ \bibnamefont
  {Braunstein}}\ and\ \bibinfo {author} {\bibfnamefont {H.~J.}\ \bibnamefont
  {Kimble}},\ }\bibfield  {title} {\bibinfo {title} {{``Teleportation of
  Continuous Quantum Variables''}},\ }\href
  {https://doi.org/10.1103/PhysRevLett.80.869} {\bibfield  {journal} {\bibinfo
  {journal} {Phys. Rev. Lett.}\ }\textbf {\bibinfo {volume} {80}},\ \bibinfo
  {pages} {869} (\bibinfo {year} {1998})}\BibitemShut {NoStop}%
\bibitem [{\citenamefont {Kimble}(2008)}]{Kimble2008}%
  \BibitemOpen
  \bibfield  {author} {\bibinfo {author} {\bibfnamefont {H.~J.}\ \bibnamefont
  {Kimble}},\ }\bibfield  {title} {\bibinfo {title} {{``The quantum
  internet''}},\ }\href {https://doi.org/10.1038/nature07127} {\bibfield
  {journal} {\bibinfo  {journal} {Nature}\ }\textbf {\bibinfo {volume} {453}},\
  \bibinfo {pages} {1023} (\bibinfo {year} {2008})}\BibitemShut {NoStop}%
\bibitem [{\citenamefont {DiVincenzo}(1995)}]{DiVincenzo1995}%
  \BibitemOpen
  \bibfield  {author} {\bibinfo {author} {\bibfnamefont {D.~P.}\ \bibnamefont
  {DiVincenzo}},\ }\bibfield  {title} {\bibinfo {title} {{``Quantum
  Computation''}},\ }\href {https://doi.org/doi:10.1126/science.270.5234.255}
  {\bibfield  {journal} {\bibinfo  {journal} {Science}\ }\textbf {\bibinfo
  {volume} {270}},\ \bibinfo {pages} {255} (\bibinfo {year}
  {1995})}\BibitemShut {NoStop}%
\bibitem [{\citenamefont {Lloyd}\ and\ \citenamefont
  {Braunstein}(1999)}]{Lloyd1999}%
  \BibitemOpen
  \bibfield  {author} {\bibinfo {author} {\bibfnamefont {S.}~\bibnamefont
  {Lloyd}}\ and\ \bibinfo {author} {\bibfnamefont {S.~L.}\ \bibnamefont
  {Braunstein}},\ }\bibfield  {title} {\bibinfo {title} {{``Quantum Computation
  over Continuous Variables''}},\ }\href
  {https://doi.org/10.1103/PhysRevLett.82.1784} {\bibfield  {journal} {\bibinfo
   {journal} {Phys. Rev. Lett.}\ }\textbf {\bibinfo {volume} {82}},\ \bibinfo
  {pages} {1784} (\bibinfo {year} {1999})}\BibitemShut {NoStop}%
\bibitem [{\citenamefont {Menicucci}\ \emph {et~al.}(2006)\citenamefont
  {Menicucci}, \citenamefont {{van Loock}}, \citenamefont {Gu}, \citenamefont
  {Weedbrook}, \citenamefont {Ralph},\ and\ \citenamefont
  {Nielsen}}]{Menicucci2006}%
  \BibitemOpen
  \bibfield  {author} {\bibinfo {author} {\bibfnamefont {N.~C.}\ \bibnamefont
  {Menicucci}}, \bibinfo {author} {\bibfnamefont {P.}~\bibnamefont {{van
  Loock}}}, \bibinfo {author} {\bibfnamefont {M.}~\bibnamefont {Gu}}, \bibinfo
  {author} {\bibfnamefont {C.}~\bibnamefont {Weedbrook}}, \bibinfo {author}
  {\bibfnamefont {T.~C.}\ \bibnamefont {Ralph}},\ and\ \bibinfo {author}
  {\bibfnamefont {M.~A.}\ \bibnamefont {Nielsen}},\ }\bibfield  {title}
  {\bibinfo {title} {{``Universal Quantum Computation with Continuous-Variable
  Cluster States''}},\ }\href {https://doi.org/10.1103/PhysRevLett.97.110501}
  {\bibfield  {journal} {\bibinfo  {journal} {Phys. Rev. Lett.}\ }\textbf
  {\bibinfo {volume} {97}},\ \bibinfo {pages} {110501} (\bibinfo {year}
  {2006})}\BibitemShut {NoStop}%
\bibitem [{\citenamefont {Wineland}\ \emph {et~al.}(1992)\citenamefont
  {Wineland}, \citenamefont {Bollinger}, \citenamefont {Itano}, \citenamefont
  {Moore},\ and\ \citenamefont {Heinzen}}]{Wineland1992}%
  \BibitemOpen
  \bibfield  {author} {\bibinfo {author} {\bibfnamefont {D.~J.}\ \bibnamefont
  {Wineland}}, \bibinfo {author} {\bibfnamefont {J.~J.}\ \bibnamefont
  {Bollinger}}, \bibinfo {author} {\bibfnamefont {W.~M.}\ \bibnamefont
  {Itano}}, \bibinfo {author} {\bibfnamefont {F.~L.}\ \bibnamefont {Moore}},\
  and\ \bibinfo {author} {\bibfnamefont {D.~J.}\ \bibnamefont {Heinzen}},\
  }\bibfield  {title} {\bibinfo {title} {{``Spin squeezing and reduced quantum
  noise in spectroscopy''}},\ }\href
  {https://doi.org/10.1103/PhysRevA.46.R6797} {\bibfield  {journal} {\bibinfo
  {journal} {Phys. Rev. A}\ }\textbf {\bibinfo {volume} {46}},\ \bibinfo
  {pages} {R6797} (\bibinfo {year} {1992})}\BibitemShut {NoStop}%
\bibitem [{\citenamefont {Huelga}\ \emph {et~al.}(1997)\citenamefont {Huelga},
  \citenamefont {Macchiavello}, \citenamefont {Pellizzari}, \citenamefont
  {Ekert}, \citenamefont {Plenio},\ and\ \citenamefont {Cirac}}]{Huelga1997}%
  \BibitemOpen
  \bibfield  {author} {\bibinfo {author} {\bibfnamefont {S.~F.}\ \bibnamefont
  {Huelga}}, \bibinfo {author} {\bibfnamefont {C.}~\bibnamefont
  {Macchiavello}}, \bibinfo {author} {\bibfnamefont {T.}~\bibnamefont
  {Pellizzari}}, \bibinfo {author} {\bibfnamefont {A.~K.}\ \bibnamefont
  {Ekert}}, \bibinfo {author} {\bibfnamefont {M.~B.}\ \bibnamefont {Plenio}},\
  and\ \bibinfo {author} {\bibfnamefont {J.~I.}\ \bibnamefont {Cirac}},\
  }\bibfield  {title} {\bibinfo {title} {{``Improvement of frequency standards
  with quantum entanglement''}},\ }\href
  {https://doi.org/10.1103/PhysRevLett.79.3865} {\bibfield  {journal} {\bibinfo
   {journal} {Phys. Rev. Lett.}\ }\textbf {\bibinfo {volume} {79}},\ \bibinfo
  {pages} {3865} (\bibinfo {year} {1997})}\BibitemShut {NoStop}%
\bibitem [{\citenamefont {Giovannetti}\ \emph {et~al.}(2011)\citenamefont
  {Giovannetti}, \citenamefont {Lloyd},\ and\ \citenamefont
  {Maccone}}]{Giovannetti2011}%
  \BibitemOpen
  \bibfield  {author} {\bibinfo {author} {\bibfnamefont {V.}~\bibnamefont
  {Giovannetti}}, \bibinfo {author} {\bibfnamefont {S.}~\bibnamefont {Lloyd}},\
  and\ \bibinfo {author} {\bibfnamefont {L.}~\bibnamefont {Maccone}},\
  }\bibfield  {title} {\bibinfo {title} {{``Advances in quantum metrology''}},\
  }\href {https://doi.org/10.1038/nphoton.2011.35} {\bibfield  {journal}
  {\bibinfo  {journal} {Nat. Photonics}\ }\textbf {\bibinfo {volume} {5}},\
  \bibinfo {pages} {222} (\bibinfo {year} {2011})}\BibitemShut {NoStop}%
\bibitem [{\citenamefont {Binnig}\ \emph {et~al.}(1986)\citenamefont {Binnig},
  \citenamefont {Quate},\ and\ \citenamefont {Gerber}}]{Binnig1986}%
  \BibitemOpen
  \bibfield  {author} {\bibinfo {author} {\bibfnamefont {G.}~\bibnamefont
  {Binnig}}, \bibinfo {author} {\bibfnamefont {C.~F.}\ \bibnamefont {Quate}},\
  and\ \bibinfo {author} {\bibfnamefont {C.}~\bibnamefont {Gerber}},\
  }\bibfield  {title} {\bibinfo {title} {{``Atomic Force Microscope''}},\
  }\href {https://doi.org/10.1103/PhysRevLett.56.930} {\bibfield  {journal}
  {\bibinfo  {journal} {Phys. Rev. Lett.}\ }\textbf {\bibinfo {volume} {56}},\
  \bibinfo {pages} {930} (\bibinfo {year} {1986})}\BibitemShut {NoStop}%
\bibitem [{\citenamefont {Chan}\ \emph {et~al.}(2001)\citenamefont {Chan},
  \citenamefont {Aksyuk}, \citenamefont {Kleiman}, \citenamefont {Bishop},\
  and\ \citenamefont {Capasso}}]{Chan2001}%
  \BibitemOpen
  \bibfield  {author} {\bibinfo {author} {\bibfnamefont {H.~B.}\ \bibnamefont
  {Chan}}, \bibinfo {author} {\bibfnamefont {V.~A.}\ \bibnamefont {Aksyuk}},
  \bibinfo {author} {\bibfnamefont {R.~N.}\ \bibnamefont {Kleiman}}, \bibinfo
  {author} {\bibfnamefont {D.~J.}\ \bibnamefont {Bishop}},\ and\ \bibinfo
  {author} {\bibfnamefont {F.}~\bibnamefont {Capasso}},\ }\bibfield  {title}
  {\bibinfo {title} {{``Quantum mechanical actuation of microelectromechanical
  systems by the Casimir force''}},\ }\href
  {https://doi.org/10.1126/science.1057984} {\bibfield  {journal} {\bibinfo
  {journal} {Science}\ }\textbf {\bibinfo {volume} {291}},\ \bibinfo {pages}
  {1941} (\bibinfo {year} {2001})}\BibitemShut {NoStop}%
\bibitem [{\citenamefont {Gavartin}\ \emph {et~al.}(2012)\citenamefont
  {Gavartin}, \citenamefont {Verlot},\ and\ \citenamefont
  {Kippenberg}}]{Gavartin2012}%
  \BibitemOpen
  \bibfield  {author} {\bibinfo {author} {\bibfnamefont {E.}~\bibnamefont
  {Gavartin}}, \bibinfo {author} {\bibfnamefont {P.}~\bibnamefont {Verlot}},\
  and\ \bibinfo {author} {\bibfnamefont {T.~J.}\ \bibnamefont {Kippenberg}},\
  }\bibfield  {title} {\bibinfo {title} {{``A hybrid on-chip optomechanical
  transducer for ultrasensitive force measurements''}},\ }\href
  {https://doi.org/10.1038/nnano.2012.97} {\bibfield  {journal} {\bibinfo
  {journal} {Nat. Nanotechnol.}\ }\textbf {\bibinfo {volume} {7}},\ \bibinfo
  {pages} {509} (\bibinfo {year} {2012})}\BibitemShut {NoStop}%
\bibitem [{\citenamefont {Moser}\ \emph {et~al.}(2013)\citenamefont {Moser},
  \citenamefont {G{\"u}ttinger}, \citenamefont {Eichler}, \citenamefont
  {Esplandiu}, \citenamefont {Liu}, \citenamefont {Dykman},\ and\ \citenamefont
  {Bachtold}}]{Moser2013}%
  \BibitemOpen
  \bibfield  {author} {\bibinfo {author} {\bibfnamefont {J.}~\bibnamefont
  {Moser}}, \bibinfo {author} {\bibfnamefont {J.}~\bibnamefont
  {G{\"u}ttinger}}, \bibinfo {author} {\bibfnamefont {A.}~\bibnamefont
  {Eichler}}, \bibinfo {author} {\bibfnamefont {M.~J.}\ \bibnamefont
  {Esplandiu}}, \bibinfo {author} {\bibfnamefont {D.~E.}\ \bibnamefont {Liu}},
  \bibinfo {author} {\bibfnamefont {M.~I.}\ \bibnamefont {Dykman}},\ and\
  \bibinfo {author} {\bibfnamefont {A.}~\bibnamefont {Bachtold}},\ }\bibfield
  {title} {\bibinfo {title} {{``Ultrasensitive force detection with a nanotube
  mechanical resonator''}},\ }\href {https://doi.org/10.1038/nnano.2013.97}
  {\bibfield  {journal} {\bibinfo  {journal} {Nat. Nanotechnol.}\ }\textbf
  {\bibinfo {volume} {8}},\ \bibinfo {pages} {493} (\bibinfo {year}
  {2013})}\BibitemShut {NoStop}%
\bibitem [{\citenamefont {Eisert}\ \emph {et~al.}(2004)\citenamefont {Eisert},
  \citenamefont {Plenio}, \citenamefont {Bose},\ and\ \citenamefont
  {Hartley}}]{Eisert2004}%
  \BibitemOpen
  \bibfield  {author} {\bibinfo {author} {\bibfnamefont {J.}~\bibnamefont
  {Eisert}}, \bibinfo {author} {\bibfnamefont {M.~B.}\ \bibnamefont {Plenio}},
  \bibinfo {author} {\bibfnamefont {S.}~\bibnamefont {Bose}},\ and\ \bibinfo
  {author} {\bibfnamefont {J.}~\bibnamefont {Hartley}},\ }\bibfield  {title}
  {\bibinfo {title} {{``Towards Quantum Entanglement in Nanoelectromechanical
  Devices''}},\ }\href {https://doi.org/10.1103/PhysRevLett.93.190402}
  {\bibfield  {journal} {\bibinfo  {journal} {Phys. Rev. Lett.}\ }\textbf
  {\bibinfo {volume} {93}},\ \bibinfo {pages} {190402} (\bibinfo {year}
  {2004})}\BibitemShut {NoStop}%
\bibitem [{\citenamefont {Plenio}\ \emph {et~al.}(2004)\citenamefont {Plenio},
  \citenamefont {Hartley},\ and\ \citenamefont {Eisert}}]{Plenio2004}%
  \BibitemOpen
  \bibfield  {author} {\bibinfo {author} {\bibfnamefont {M.~B.}\ \bibnamefont
  {Plenio}}, \bibinfo {author} {\bibfnamefont {J.}~\bibnamefont {Hartley}},\
  and\ \bibinfo {author} {\bibfnamefont {J.}~\bibnamefont {Eisert}},\
  }\bibfield  {title} {\bibinfo {title} {{``Dynamics and manipulation of
  entanglement in coupled harmonic systems with many degrees of freedom''}},\
  }\href {https://doi.org/10.1088/1367-2630/6/1/036} {\bibfield  {journal}
  {\bibinfo  {journal} {New J. Phys.}\ }\textbf {\bibinfo {volume} {6}},\
  \bibinfo {pages} {36} (\bibinfo {year} {2004})}\BibitemShut {NoStop}%
\bibitem [{\citenamefont {Pirandola}\ \emph {et~al.}(2006)\citenamefont
  {Pirandola}, \citenamefont {Vitali}, \citenamefont {Tombesi},\ and\
  \citenamefont {Lloyd}}]{Pirandola2006}%
  \BibitemOpen
  \bibfield  {author} {\bibinfo {author} {\bibfnamefont {S.}~\bibnamefont
  {Pirandola}}, \bibinfo {author} {\bibfnamefont {D.}~\bibnamefont {Vitali}},
  \bibinfo {author} {\bibfnamefont {P.}~\bibnamefont {Tombesi}},\ and\ \bibinfo
  {author} {\bibfnamefont {S.}~\bibnamefont {Lloyd}},\ }\bibfield  {title}
  {\bibinfo {title} {{``Macroscopic Entanglement by Entanglement Swapping''}},\
  }\href {https://doi.org/10.1103/PhysRevLett.97.150403} {\bibfield  {journal}
  {\bibinfo  {journal} {Phys. Rev. Lett.}\ }\textbf {\bibinfo {volume} {97}},\
  \bibinfo {pages} {150403} (\bibinfo {year} {2006})}\BibitemShut {NoStop}%
\bibitem [{\citenamefont {Jost}\ \emph {et~al.}(2009)\citenamefont {Jost},
  \citenamefont {Home}, \citenamefont {Amini}, \citenamefont {Hanneke},
  \citenamefont {Ozeri}, \citenamefont {Langer}, \citenamefont {Bollinger},
  \citenamefont {Leibfried},\ and\ \citenamefont {Wineland}}]{Jost2009}%
  \BibitemOpen
  \bibfield  {author} {\bibinfo {author} {\bibfnamefont {J.~D.}\ \bibnamefont
  {Jost}}, \bibinfo {author} {\bibfnamefont {J.~P.}\ \bibnamefont {Home}},
  \bibinfo {author} {\bibfnamefont {J.~M.}\ \bibnamefont {Amini}}, \bibinfo
  {author} {\bibfnamefont {D.}~\bibnamefont {Hanneke}}, \bibinfo {author}
  {\bibfnamefont {R.}~\bibnamefont {Ozeri}}, \bibinfo {author} {\bibfnamefont
  {C.}~\bibnamefont {Langer}}, \bibinfo {author} {\bibfnamefont {J.~J.}\
  \bibnamefont {Bollinger}}, \bibinfo {author} {\bibfnamefont {D.}~\bibnamefont
  {Leibfried}},\ and\ \bibinfo {author} {\bibfnamefont {D.~J.}\ \bibnamefont
  {Wineland}},\ }\bibfield  {title} {\bibinfo {title} {{``Entangled mechanical
  oscillators''}},\ }\href {https://doi.org/10.1038/nature08006} {\bibfield
  {journal} {\bibinfo  {journal} {Nature}\ }\textbf {\bibinfo {volume} {459}},\
  \bibinfo {pages} {683} (\bibinfo {year} {2009})}\BibitemShut {NoStop}%
\bibitem [{\citenamefont {Abdi}\ \emph {et~al.}(2012)\citenamefont {Abdi},
  \citenamefont {Pirandola}, \citenamefont {Tombesi},\ and\ \citenamefont
  {Vitali}}]{Abdi2012}%
  \BibitemOpen
  \bibfield  {author} {\bibinfo {author} {\bibfnamefont {M.}~\bibnamefont
  {Abdi}}, \bibinfo {author} {\bibfnamefont {S.}~\bibnamefont {Pirandola}},
  \bibinfo {author} {\bibfnamefont {P.}~\bibnamefont {Tombesi}},\ and\ \bibinfo
  {author} {\bibfnamefont {D.}~\bibnamefont {Vitali}},\ }\bibfield  {title}
  {\bibinfo {title} {{``Entanglement Swapping with Local Certification:
  Application to Remote Micromechanical Resonators''}},\ }\href
  {https://doi.org/10.1103/PhysRevLett.109.143601} {\bibfield  {journal}
  {\bibinfo  {journal} {Phys. Rev. Lett.}\ }\textbf {\bibinfo {volume} {109}},\
  \bibinfo {pages} {143601} (\bibinfo {year} {2012})}\BibitemShut {NoStop}%
\bibitem [{\citenamefont {Ockeloen-Korppi}\ \emph {et~al.}(2018)\citenamefont
  {Ockeloen-Korppi}, \citenamefont {Damsk{\"a}gg}, \citenamefont
  {Pirkkalainen}, \citenamefont {Asjad}, \citenamefont {Clerk}, \citenamefont
  {Massel}, \citenamefont {Woolley},\ and\ \citenamefont
  {Sillanp{\"a}{\"a}}}]{Ockeloen2018}%
  \BibitemOpen
  \bibfield  {author} {\bibinfo {author} {\bibfnamefont {C.~F.}\ \bibnamefont
  {Ockeloen-Korppi}}, \bibinfo {author} {\bibfnamefont {E.}~\bibnamefont
  {Damsk{\"a}gg}}, \bibinfo {author} {\bibfnamefont {J.-M.}\ \bibnamefont
  {Pirkkalainen}}, \bibinfo {author} {\bibfnamefont {M.}~\bibnamefont {Asjad}},
  \bibinfo {author} {\bibfnamefont {A.~A.}\ \bibnamefont {Clerk}}, \bibinfo
  {author} {\bibfnamefont {F.}~\bibnamefont {Massel}}, \bibinfo {author}
  {\bibfnamefont {M.~J.}\ \bibnamefont {Woolley}},\ and\ \bibinfo {author}
  {\bibfnamefont {M.~A.}\ \bibnamefont {Sillanp{\"a}{\"a}}},\ }\bibfield
  {title} {\bibinfo {title} {{``Stabilized entanglement of massive mechanical
  oscillators''}},\ }\href {https://doi.org/10.1038/s41586-018-0038-x}
  {\bibfield  {journal} {\bibinfo  {journal} {Nature}\ }\textbf {\bibinfo
  {volume} {556}},\ \bibinfo {pages} {478} (\bibinfo {year}
  {2018})}\BibitemShut {NoStop}%
\bibitem [{\citenamefont {Riedinger}\ \emph {et~al.}(2018)\citenamefont
  {Riedinger}, \citenamefont {Wallucks}, \citenamefont {Marinkovi{\'c}},
  \citenamefont {L{\"o}schnauer}, \citenamefont {Aspelmeyer}, \citenamefont
  {Hong},\ and\ \citenamefont {Gr{\"o}blacher}}]{Riedinger2018}%
  \BibitemOpen
  \bibfield  {author} {\bibinfo {author} {\bibfnamefont {R.}~\bibnamefont
  {Riedinger}}, \bibinfo {author} {\bibfnamefont {A.}~\bibnamefont {Wallucks}},
  \bibinfo {author} {\bibfnamefont {I.}~\bibnamefont {Marinkovi{\'c}}},
  \bibinfo {author} {\bibfnamefont {C.}~\bibnamefont {L{\"o}schnauer}},
  \bibinfo {author} {\bibfnamefont {M.}~\bibnamefont {Aspelmeyer}}, \bibinfo
  {author} {\bibfnamefont {S.}~\bibnamefont {Hong}},\ and\ \bibinfo {author}
  {\bibfnamefont {S.}~\bibnamefont {Gr{\"o}blacher}},\ }\bibfield  {title}
  {\bibinfo {title} {{``Remote quantum entanglement between two micromechanical
  oscillators''}},\ }\href {https://doi.org/10.1038/s41586-018-0036-z}
  {\bibfield  {journal} {\bibinfo  {journal} {Nature}\ }\textbf {\bibinfo
  {volume} {556}},\ \bibinfo {pages} {473} (\bibinfo {year}
  {2018})}\BibitemShut {NoStop}%
\bibitem [{\citenamefont {Schmidt}\ \emph {et~al.}(2012)\citenamefont
  {Schmidt}, \citenamefont {Ludwig},\ and\ \citenamefont
  {Marquardt}}]{Schmidt2012}%
  \BibitemOpen
  \bibfield  {author} {\bibinfo {author} {\bibfnamefont {M.}~\bibnamefont
  {Schmidt}}, \bibinfo {author} {\bibfnamefont {M.}~\bibnamefont {Ludwig}},\
  and\ \bibinfo {author} {\bibfnamefont {F.}~\bibnamefont {Marquardt}},\
  }\bibfield  {title} {\bibinfo {title} {{``Optomechanical circuits for
  nanomechanical continuous variable quantum state processing''}},\ }\href
  {https://doi.org/10.1088/1367-2630/14/12/125005} {\bibfield  {journal}
  {\bibinfo  {journal} {New J. Phys.}\ }\textbf {\bibinfo {volume} {14}},\
  \bibinfo {pages} {125005} (\bibinfo {year} {2012})}\BibitemShut {NoStop}%
\bibitem [{\citenamefont {Rips}\ and\ \citenamefont
  {Hartmann}(2013)}]{Rips2013}%
  \BibitemOpen
  \bibfield  {author} {\bibinfo {author} {\bibfnamefont {S.}~\bibnamefont
  {Rips}}\ and\ \bibinfo {author} {\bibfnamefont {M.~J.}\ \bibnamefont
  {Hartmann}},\ }\bibfield  {title} {\bibinfo {title} {{``Quantum Information
  Processing with Nanomechanical Qubits''}},\ }\href
  {https://doi.org/10.1103/PhysRevLett.110.120503} {\bibfield  {journal}
  {\bibinfo  {journal} {Phys. Rev. Lett.}\ }\textbf {\bibinfo {volume} {110}},\
  \bibinfo {pages} {120503} (\bibinfo {year} {2013})}\BibitemShut {NoStop}%
\bibitem [{\citenamefont {Zippilli}\ and\ \citenamefont
  {Vitali}(2021)}]{Zippilli2021}%
  \BibitemOpen
  \bibfield  {author} {\bibinfo {author} {\bibfnamefont {S.}~\bibnamefont
  {Zippilli}}\ and\ \bibinfo {author} {\bibfnamefont {D.}~\bibnamefont
  {Vitali}},\ }\bibfield  {title} {\bibinfo {title} {{``Dissipative Engineering
  of Gaussian Entangled States in Harmonic Lattices with a Single-Site Squeezed
  Reservoir''}},\ }\href {https://doi.org/10.1103/physrevlett.126.020402}
  {\bibfield  {journal} {\bibinfo  {journal} {Phys. Rev. Lett.}\ }\textbf
  {\bibinfo {volume} {126}},\ \bibinfo {pages} {020402} (\bibinfo {year}
  {2021})}\BibitemShut {NoStop}%
\bibitem [{\citenamefont {Vitali}\ \emph {et~al.}(2007)\citenamefont {Vitali},
  \citenamefont {Gigan}, \citenamefont {Ferreira}, \citenamefont {B{\"o}hm},
  \citenamefont {Tombesi}, \citenamefont {Guerreiro}, \citenamefont {Vedral},
  \citenamefont {Zeilinger},\ and\ \citenamefont {Aspelmeyer}}]{Vitali2007}%
  \BibitemOpen
  \bibfield  {author} {\bibinfo {author} {\bibfnamefont {D.}~\bibnamefont
  {Vitali}}, \bibinfo {author} {\bibfnamefont {S.}~\bibnamefont {Gigan}},
  \bibinfo {author} {\bibfnamefont {A.}~\bibnamefont {Ferreira}}, \bibinfo
  {author} {\bibfnamefont {H.~R.}\ \bibnamefont {B{\"o}hm}}, \bibinfo {author}
  {\bibfnamefont {P.}~\bibnamefont {Tombesi}}, \bibinfo {author} {\bibfnamefont
  {A.}~\bibnamefont {Guerreiro}}, \bibinfo {author} {\bibfnamefont
  {V.}~\bibnamefont {Vedral}}, \bibinfo {author} {\bibfnamefont
  {A.}~\bibnamefont {Zeilinger}},\ and\ \bibinfo {author} {\bibfnamefont
  {M.}~\bibnamefont {Aspelmeyer}},\ }\bibfield  {title} {\bibinfo {title}
  {{``Optomechanical Entanglement between a Movable Mirror and a Cavity
  Field''}},\ }\href {https://doi.org/10.1103/PhysRevLett.98.030405} {\bibfield
   {journal} {\bibinfo  {journal} {Phys. Rev. Lett.}\ }\textbf {\bibinfo
  {volume} {98}},\ \bibinfo {pages} {030405} (\bibinfo {year}
  {2007})}\BibitemShut {NoStop}%
\bibitem [{\citenamefont {Paternostro}\ \emph {et~al.}(2007)\citenamefont
  {Paternostro}, \citenamefont {Vitali}, \citenamefont {Gigan}, \citenamefont
  {Kim}, \citenamefont {Brukner}, \citenamefont {Eisert},\ and\ \citenamefont
  {Aspelmeyer}}]{Paternostro2007}%
  \BibitemOpen
  \bibfield  {author} {\bibinfo {author} {\bibfnamefont {M.}~\bibnamefont
  {Paternostro}}, \bibinfo {author} {\bibfnamefont {D.}~\bibnamefont {Vitali}},
  \bibinfo {author} {\bibfnamefont {S.}~\bibnamefont {Gigan}}, \bibinfo
  {author} {\bibfnamefont {M.~S.}\ \bibnamefont {Kim}}, \bibinfo {author}
  {\bibfnamefont {C.}~\bibnamefont {Brukner}}, \bibinfo {author} {\bibfnamefont
  {J.}~\bibnamefont {Eisert}},\ and\ \bibinfo {author} {\bibfnamefont
  {M.}~\bibnamefont {Aspelmeyer}},\ }\bibfield  {title} {\bibinfo {title}
  {{``Creating and Probing Multipartite Macroscopic Entanglement with
  Light''}},\ }\href {https://doi.org/10.1103/PhysRevLett.99.250401} {\bibfield
   {journal} {\bibinfo  {journal} {Phys. Rev. Lett.}\ }\textbf {\bibinfo
  {volume} {99}},\ \bibinfo {pages} {250401} (\bibinfo {year}
  {2007})}\BibitemShut {NoStop}%
\bibitem [{\citenamefont {Favero}\ and\ \citenamefont
  {Karrai}(2009)}]{Favero2009}%
  \BibitemOpen
  \bibfield  {author} {\bibinfo {author} {\bibfnamefont {I.}~\bibnamefont
  {Favero}}\ and\ \bibinfo {author} {\bibfnamefont {K.}~\bibnamefont
  {Karrai}},\ }\bibfield  {title} {\bibinfo {title} {{``Optomechanics of
  deformable optical cavities''}},\ }\href
  {https://doi.org/10.1038/nphoton.2009.42} {\bibfield  {journal} {\bibinfo
  {journal} {Nat. Photonics}\ }\textbf {\bibinfo {volume} {3}},\ \bibinfo
  {pages} {201} (\bibinfo {year} {2009})}\BibitemShut {NoStop}%
\bibitem [{\citenamefont {Blencowe}(2004)}]{Blencowe2004}%
  \BibitemOpen
  \bibfield  {author} {\bibinfo {author} {\bibfnamefont {M.}~\bibnamefont
  {Blencowe}},\ }\bibfield  {title} {\bibinfo {title} {{``Quantum
  electromechanical systems''}},\ }\href
  {https://doi.org/10.1016/j.physrep.2003.12.005} {\bibfield  {journal}
  {\bibinfo  {journal} {Phys. Rep.}\ }\textbf {\bibinfo {volume} {395}},\
  \bibinfo {pages} {159} (\bibinfo {year} {2004})}\BibitemShut {NoStop}%
\bibitem [{\citenamefont {Abdi}\ \emph {et~al.}(2015)\citenamefont {Abdi},
  \citenamefont {Pernpeintner}, \citenamefont {Gross}, \citenamefont {Huebl},\
  and\ \citenamefont {Hartmann}}]{Abdi2015a}%
  \BibitemOpen
  \bibfield  {author} {\bibinfo {author} {\bibfnamefont {M.}~\bibnamefont
  {Abdi}}, \bibinfo {author} {\bibfnamefont {M.}~\bibnamefont {Pernpeintner}},
  \bibinfo {author} {\bibfnamefont {R.}~\bibnamefont {Gross}}, \bibinfo
  {author} {\bibfnamefont {H.}~\bibnamefont {Huebl}},\ and\ \bibinfo {author}
  {\bibfnamefont {M.~J.}\ \bibnamefont {Hartmann}},\ }\bibfield  {title}
  {\bibinfo {title} {{``Quantum State Engineering with Circuit
  Electromechanical Three-Body Interactions''}},\ }\href
  {https://doi.org/10.1103/physrevlett.114.173602} {\bibfield  {journal}
  {\bibinfo  {journal} {Phys. Rev. Lett.}\ }\textbf {\bibinfo {volume} {114}},\
  \bibinfo {pages} {173602} (\bibinfo {year} {2015})}\BibitemShut {NoStop}%
\bibitem [{\citenamefont {Lee}\ \emph {et~al.}(2017)\citenamefont {Lee},
  \citenamefont {Lee}, \citenamefont {Cady}, \citenamefont {Ovartchaiyapong},\
  and\ \citenamefont {Jayich}}]{Lee2017}%
  \BibitemOpen
  \bibfield  {author} {\bibinfo {author} {\bibfnamefont {D.}~\bibnamefont
  {Lee}}, \bibinfo {author} {\bibfnamefont {K.~W.}\ \bibnamefont {Lee}},
  \bibinfo {author} {\bibfnamefont {J.~V.}\ \bibnamefont {Cady}}, \bibinfo
  {author} {\bibfnamefont {P.}~\bibnamefont {Ovartchaiyapong}},\ and\ \bibinfo
  {author} {\bibfnamefont {A.~C.~B.}\ \bibnamefont {Jayich}},\ }\bibfield
  {title} {\bibinfo {title} {{``Topical review: spins and mechanics in
  diamond''}},\ }\href {https://doi.org/10.1088/2040-8986/aa52cd} {\bibfield
  {journal} {\bibinfo  {journal} {J. Opt.}\ }\textbf {\bibinfo {volume} {19}},\
  \bibinfo {pages} {033001} (\bibinfo {year} {2017})}\BibitemShut {NoStop}%
\bibitem [{\citenamefont {Hartmann}\ and\ \citenamefont
  {Plenio}(2008)}]{Hartmann2008}%
  \BibitemOpen
  \bibfield  {author} {\bibinfo {author} {\bibfnamefont {M.~J.}\ \bibnamefont
  {Hartmann}}\ and\ \bibinfo {author} {\bibfnamefont {M.~B.}\ \bibnamefont
  {Plenio}},\ }\bibfield  {title} {\bibinfo {title} {{``Steady State
  Entanglement in the Mechanical Vibrations of Two Dielectric Membranes''}},\
  }\href {https://doi.org/10.1103/PhysRevLett.101.200503} {\bibfield  {journal}
  {\bibinfo  {journal} {Phys. Rev. Lett.}\ }\textbf {\bibinfo {volume} {101}},\
  \bibinfo {pages} {200503} (\bibinfo {year} {2008})}\BibitemShut {NoStop}%
\bibitem [{\citenamefont {Rabl}\ \emph {et~al.}(2010)\citenamefont {Rabl},
  \citenamefont {Kolkowitz}, \citenamefont {Koppens}, \citenamefont {Harris},
  \citenamefont {Zoller},\ and\ \citenamefont {Lukin}}]{Rabl2010b}%
  \BibitemOpen
  \bibfield  {author} {\bibinfo {author} {\bibfnamefont {P.}~\bibnamefont
  {Rabl}}, \bibinfo {author} {\bibfnamefont {S.~J.}\ \bibnamefont {Kolkowitz}},
  \bibinfo {author} {\bibfnamefont {F.~H.~L.}\ \bibnamefont {Koppens}},
  \bibinfo {author} {\bibfnamefont {J.~G.~E.}\ \bibnamefont {Harris}}, \bibinfo
  {author} {\bibfnamefont {P.}~\bibnamefont {Zoller}},\ and\ \bibinfo {author}
  {\bibfnamefont {M.~D.}\ \bibnamefont {Lukin}},\ }\bibfield  {title} {\bibinfo
  {title} {{``A quantum spin transducer based on nanoelectromechanical
  resonator arrays''}},\ }\href {https://doi.org/10.1038/nphys1679} {\bibfield
  {journal} {\bibinfo  {journal} {Nat. Phys.}\ }\textbf {\bibinfo {volume}
  {6}},\ \bibinfo {pages} {602} (\bibinfo {year} {2010})}\BibitemShut {NoStop}%
\bibitem [{\citenamefont {Stannigel}\ \emph {et~al.}(2010)\citenamefont
  {Stannigel}, \citenamefont {Rabl}, \citenamefont {S{\o}rensen}, \citenamefont
  {Zoller},\ and\ \citenamefont {Lukin}}]{Stannigel2010}%
  \BibitemOpen
  \bibfield  {author} {\bibinfo {author} {\bibfnamefont {K.}~\bibnamefont
  {Stannigel}}, \bibinfo {author} {\bibfnamefont {P.}~\bibnamefont {Rabl}},
  \bibinfo {author} {\bibfnamefont {A.~S.}\ \bibnamefont {S{\o}rensen}},
  \bibinfo {author} {\bibfnamefont {P.}~\bibnamefont {Zoller}},\ and\ \bibinfo
  {author} {\bibfnamefont {M.~D.}\ \bibnamefont {Lukin}},\ }\bibfield  {title}
  {\bibinfo {title} {{``Optomechanical Transducers for Long-Distance Quantum
  Communication''}},\ }\href {https://doi.org/10.1103/PhysRevLett.105.220501}
  {\bibfield  {journal} {\bibinfo  {journal} {Phys. Rev. Lett.}\ }\textbf
  {\bibinfo {volume} {105}},\ \bibinfo {pages} {220501} (\bibinfo {year}
  {2010})}\BibitemShut {NoStop}%
\bibitem [{\citenamefont {Abdi}\ and\ \citenamefont
  {Hartmann}(2015)}]{Abdi2015b}%
  \BibitemOpen
  \bibfield  {author} {\bibinfo {author} {\bibfnamefont {M.}~\bibnamefont
  {Abdi}}\ and\ \bibinfo {author} {\bibfnamefont {M.~J.}\ \bibnamefont
  {Hartmann}},\ }\bibfield  {title} {\bibinfo {title} {{``Entangling the motion
  of two optically trapped objects via time- modulated driving fields''}},\
  }\href {https://doi.org/10.1088/1367-2630/17/1/013056} {\bibfield  {journal}
  {\bibinfo  {journal} {New J. Phys.}\ }\textbf {\bibinfo {volume} {17}},\
  \bibinfo {pages} {013056} (\bibinfo {year} {2015})}\BibitemShut {NoStop}%
\bibitem [{\citenamefont {MacQuarrie}\ \emph {et~al.}(2013)\citenamefont
  {MacQuarrie}, \citenamefont {Gosavi}, \citenamefont {Jungwirth},
  \citenamefont {Bhave},\ and\ \citenamefont {Fuchs}}]{MacQuarrie2013}%
  \BibitemOpen
  \bibfield  {author} {\bibinfo {author} {\bibfnamefont {E.~R.}\ \bibnamefont
  {MacQuarrie}}, \bibinfo {author} {\bibfnamefont {T.~A.}\ \bibnamefont
  {Gosavi}}, \bibinfo {author} {\bibfnamefont {N.~R.}\ \bibnamefont
  {Jungwirth}}, \bibinfo {author} {\bibfnamefont {S.~A.}\ \bibnamefont
  {Bhave}},\ and\ \bibinfo {author} {\bibfnamefont {G.~D.}\ \bibnamefont
  {Fuchs}},\ }\bibfield  {title} {\bibinfo {title} {{``Mechanical Spin Control
  of Nitrogen-Vacancy Centers in Diamond''}},\ }\href
  {https://doi.org/10.1103/PhysRevLett.111.227602} {\bibfield  {journal}
  {\bibinfo  {journal} {Phys. Rev. Lett.}\ }\textbf {\bibinfo {volume} {111}},\
  \bibinfo {pages} {227602} (\bibinfo {year} {2013})}\BibitemShut {NoStop}%
\bibitem [{\citenamefont {Barfuss}\ \emph {et~al.}(2015)\citenamefont
  {Barfuss}, \citenamefont {Teissier}, \citenamefont {Neu}, \citenamefont
  {Nunnenkamp},\ and\ \citenamefont {Maletinsky}}]{Barfuss2015}%
  \BibitemOpen
  \bibfield  {author} {\bibinfo {author} {\bibfnamefont {A.}~\bibnamefont
  {Barfuss}}, \bibinfo {author} {\bibfnamefont {J.}~\bibnamefont {Teissier}},
  \bibinfo {author} {\bibfnamefont {E.}~\bibnamefont {Neu}}, \bibinfo {author}
  {\bibfnamefont {A.}~\bibnamefont {Nunnenkamp}},\ and\ \bibinfo {author}
  {\bibfnamefont {P.}~\bibnamefont {Maletinsky}},\ }\bibfield  {title}
  {\bibinfo {title} {{``Strong mechanical driving of a single electron
  spin''}},\ }\href {https://doi.org/10.1038/nphys3411} {\bibfield  {journal}
  {\bibinfo  {journal} {Nat. Phys.}\ }\textbf {\bibinfo {volume} {11}},\
  \bibinfo {pages} {820} (\bibinfo {year} {2015})}\BibitemShut {NoStop}%
\bibitem [{\citenamefont {Schuetz}\ \emph {et~al.}(2015)\citenamefont
  {Schuetz}, \citenamefont {Kessler}, \citenamefont {Giedke}, \citenamefont
  {Vandersypen}, \citenamefont {Lukin},\ and\ \citenamefont
  {Cirac}}]{Schuetz2015}%
  \BibitemOpen
  \bibfield  {author} {\bibinfo {author} {\bibfnamefont {M.~J.~A.}\
  \bibnamefont {Schuetz}}, \bibinfo {author} {\bibfnamefont {E.~M.}\
  \bibnamefont {Kessler}}, \bibinfo {author} {\bibfnamefont {G.}~\bibnamefont
  {Giedke}}, \bibinfo {author} {\bibfnamefont {L.~M.~K.}\ \bibnamefont
  {Vandersypen}}, \bibinfo {author} {\bibfnamefont {M.~D.}\ \bibnamefont
  {Lukin}},\ and\ \bibinfo {author} {\bibfnamefont {J.~I.}\ \bibnamefont
  {Cirac}},\ }\bibfield  {title} {\bibinfo {title} {{``Universal Quantum
  Transducers Based on Surface Acoustic Waves''}},\ }\href
  {https://doi.org/10.1103/PhysRevX.5.031031} {\bibfield  {journal} {\bibinfo
  {journal} {Phys. Rev. X}\ }\textbf {\bibinfo {volume} {5}},\ \bibinfo {pages}
  {031031} (\bibinfo {year} {2015})}\BibitemShut {NoStop}%
\bibitem [{\citenamefont {Maity}\ \emph {et~al.}(2020)\citenamefont {Maity},
  \citenamefont {Shao}, \citenamefont {Bogdanovi{\'c}}, \citenamefont
  {Meesala}, \citenamefont {Sohn}, \citenamefont {Sinclair}, \citenamefont
  {Pingault}, \citenamefont {Chalupnik}, \citenamefont {Chia}, \citenamefont
  {Zheng}, \citenamefont {Lai},\ and\ \citenamefont {Lon{\v c}ar}}]{Maity2020}%
  \BibitemOpen
  \bibfield  {author} {\bibinfo {author} {\bibfnamefont {S.}~\bibnamefont
  {Maity}}, \bibinfo {author} {\bibfnamefont {L.}~\bibnamefont {Shao}},
  \bibinfo {author} {\bibfnamefont {S.}~\bibnamefont {Bogdanovi{\'c}}},
  \bibinfo {author} {\bibfnamefont {S.}~\bibnamefont {Meesala}}, \bibinfo
  {author} {\bibfnamefont {Y.-I.}\ \bibnamefont {Sohn}}, \bibinfo {author}
  {\bibfnamefont {N.}~\bibnamefont {Sinclair}}, \bibinfo {author}
  {\bibfnamefont {B.}~\bibnamefont {Pingault}}, \bibinfo {author}
  {\bibfnamefont {M.}~\bibnamefont {Chalupnik}}, \bibinfo {author}
  {\bibfnamefont {C.}~\bibnamefont {Chia}}, \bibinfo {author} {\bibfnamefont
  {L.}~\bibnamefont {Zheng}}, \bibinfo {author} {\bibfnamefont
  {K.}~\bibnamefont {Lai}},\ and\ \bibinfo {author} {\bibfnamefont
  {M.}~\bibnamefont {Lon{\v c}ar}},\ }\bibfield  {title} {\bibinfo {title}
  {{``Coherent acoustic control of a single silicon vacancy spin in
  diamond''}},\ }\href {https://doi.org/10.1038/s41467-019-13822-x} {\bibfield
  {journal} {\bibinfo  {journal} {Nat. Commun.}\ }\textbf {\bibinfo {volume}
  {11}},\ \bibinfo {pages} {193} (\bibinfo {year} {2020})}\BibitemShut
  {NoStop}%
\bibitem [{\citenamefont {Bennett}\ \emph {et~al.}(2013)\citenamefont
  {Bennett}, \citenamefont {Yao}, \citenamefont {Otterbach}, \citenamefont
  {Zoller}, \citenamefont {Rabl},\ and\ \citenamefont {Lukin}}]{Bennett2013}%
  \BibitemOpen
  \bibfield  {author} {\bibinfo {author} {\bibfnamefont {S.~D.}\ \bibnamefont
  {Bennett}}, \bibinfo {author} {\bibfnamefont {N.~Y.}\ \bibnamefont {Yao}},
  \bibinfo {author} {\bibfnamefont {J.}~\bibnamefont {Otterbach}}, \bibinfo
  {author} {\bibfnamefont {P.}~\bibnamefont {Zoller}}, \bibinfo {author}
  {\bibfnamefont {P.}~\bibnamefont {Rabl}},\ and\ \bibinfo {author}
  {\bibfnamefont {M.~D.}\ \bibnamefont {Lukin}},\ }\bibfield  {title} {\bibinfo
  {title} {{``Phonon-Induced Spin-Spin Interactions in Diamond Nanostructures:
  Application to Spin Squeezing''}},\ }\href
  {https://doi.org/10.1103/PhysRevLett.110.156402} {\bibfield  {journal}
  {\bibinfo  {journal} {Phys. Rev. Lett.}\ }\textbf {\bibinfo {volume} {110}},\
  \bibinfo {pages} {156402} (\bibinfo {year} {2013})}\BibitemShut {NoStop}%
\bibitem [{\citenamefont {Albrecht}\ \emph {et~al.}(2013)\citenamefont
  {Albrecht}, \citenamefont {Retzker}, \citenamefont {Jelezko},\ and\
  \citenamefont {Plenio}}]{Albrecht2013}%
  \BibitemOpen
  \bibfield  {author} {\bibinfo {author} {\bibfnamefont {A.}~\bibnamefont
  {Albrecht}}, \bibinfo {author} {\bibfnamefont {A.}~\bibnamefont {Retzker}},
  \bibinfo {author} {\bibfnamefont {F.}~\bibnamefont {Jelezko}},\ and\ \bibinfo
  {author} {\bibfnamefont {M.~B.}\ \bibnamefont {Plenio}},\ }\bibfield  {title}
  {\bibinfo {title} {{``Coupling of nitrogen vacancy centres in nanodiamonds by
  means of phonons''}},\ }\href {https://doi.org/10.1088/1367-2630/15/8/083014}
  {\bibfield  {journal} {\bibinfo  {journal} {New J. Phys.}\ }\textbf {\bibinfo
  {volume} {15}},\ \bibinfo {pages} {083014} (\bibinfo {year}
  {2013})}\BibitemShut {NoStop}%
\bibitem [{\citenamefont {Lemonde}\ \emph {et~al.}(2018)\citenamefont
  {Lemonde}, \citenamefont {Meesala}, \citenamefont {Sipahigil}, \citenamefont
  {Schuetz}, \citenamefont {Lukin}, \citenamefont {Loncar},\ and\ \citenamefont
  {Rabl}}]{Lemonde2018}%
  \BibitemOpen
  \bibfield  {author} {\bibinfo {author} {\bibfnamefont {M.-A.}\ \bibnamefont
  {Lemonde}}, \bibinfo {author} {\bibfnamefont {S.}~\bibnamefont {Meesala}},
  \bibinfo {author} {\bibfnamefont {A.}~\bibnamefont {Sipahigil}}, \bibinfo
  {author} {\bibfnamefont {M.~J.~A.}\ \bibnamefont {Schuetz}}, \bibinfo
  {author} {\bibfnamefont {M.~D.}\ \bibnamefont {Lukin}}, \bibinfo {author}
  {\bibfnamefont {M.}~\bibnamefont {Loncar}},\ and\ \bibinfo {author}
  {\bibfnamefont {P.}~\bibnamefont {Rabl}},\ }\bibfield  {title} {\bibinfo
  {title} {{``Phonon Networks with Silicon-Vacancy Centers in Diamond
  Waveguides''}},\ }\href {https://doi.org/10.1103/PhysRevLett.120.213603}
  {\bibfield  {journal} {\bibinfo  {journal} {Phys. Rev. Lett.}\ }\textbf
  {\bibinfo {volume} {120}},\ \bibinfo {pages} {213603} (\bibinfo {year}
  {2018})}\BibitemShut {NoStop}%
\bibitem [{\citenamefont {Safavi-Naeini}\ \emph {et~al.}(2019)\citenamefont
  {Safavi-Naeini}, \citenamefont {Thourhout}, \citenamefont {Baets},\ and\
  \citenamefont {Laer}}]{Safavi2019}%
  \BibitemOpen
  \bibfield  {author} {\bibinfo {author} {\bibfnamefont {A.~H.}\ \bibnamefont
  {Safavi-Naeini}}, \bibinfo {author} {\bibfnamefont {D.~V.}\ \bibnamefont
  {Thourhout}}, \bibinfo {author} {\bibfnamefont {R.}~\bibnamefont {Baets}},\
  and\ \bibinfo {author} {\bibfnamefont {R.~V.}\ \bibnamefont {Laer}},\
  }\bibfield  {title} {\bibinfo {title} {{``Controlling phonons and photons at
  the wavelength scale: integrated photonics meets integrated phononics''}},\
  }\href {https://doi.org/10.1364/OPTICA.6.000213} {\bibfield  {journal}
  {\bibinfo  {journal} {Optica}\ }\textbf {\bibinfo {volume} {6}},\ \bibinfo
  {pages} {213} (\bibinfo {year} {2019})}\BibitemShut {NoStop}%
\bibitem [{\citenamefont {Xuereb}\ \emph {et~al.}(2012)\citenamefont {Xuereb},
  \citenamefont {Genes}, ,\ and\ \citenamefont {Dantan}}]{Xuereb2012}%
  \BibitemOpen
  \bibfield  {author} {\bibinfo {author} {\bibfnamefont {A.}~\bibnamefont
  {Xuereb}}, \bibinfo {author} {\bibfnamefont {C.}~\bibnamefont {Genes}}, ,\
  and\ \bibinfo {author} {\bibfnamefont {A.}~\bibnamefont {Dantan}},\
  }\bibfield  {title} {\bibinfo {title} {{``Strong Coupling and Long-Range
  Collective Interactions in Optomechanical Arrays''}},\ }\href
  {https://doi.org/10.1103/PhysRevLett.109.223601} {\bibfield  {journal}
  {\bibinfo  {journal} {Phys. Rev. Lett.}\ }\textbf {\bibinfo {volume} {109}},\
  \bibinfo {pages} {223601} (\bibinfo {year} {2012})}\BibitemShut {NoStop}%
\bibitem [{\citenamefont {Xu}\ \emph {et~al.}(2013)\citenamefont {Xu},
  \citenamefont {Zhao},\ and\ \citenamefont {Liu}}]{Xu2013}%
  \BibitemOpen
  \bibfield  {author} {\bibinfo {author} {\bibfnamefont {X.-W.}\ \bibnamefont
  {Xu}}, \bibinfo {author} {\bibfnamefont {Y.-J.}\ \bibnamefont {Zhao}},\ and\
  \bibinfo {author} {\bibfnamefont {Y.-X.}\ \bibnamefont {Liu}},\ }\bibfield
  {title} {\bibinfo {title} {{``Entangled-state engineering of vibrational
  modes in a multimembrane optomechanical system''}},\ }\href
  {https://doi.org/10.1103/PhysRevA.88.022325} {\bibfield  {journal} {\bibinfo
  {journal} {Phys. Rev. A}\ }\textbf {\bibinfo {volume} {88}},\ \bibinfo
  {pages} {022325} (\bibinfo {year} {2013})}\BibitemShut {NoStop}%
\bibitem [{\citenamefont {Seok}\ \emph {et~al.}(2013)\citenamefont {Seok},
  \citenamefont {Buchmann}, \citenamefont {Wright},\ and\ \citenamefont
  {Meystre}}]{Seok2013}%
  \BibitemOpen
  \bibfield  {author} {\bibinfo {author} {\bibfnamefont {H.}~\bibnamefont
  {Seok}}, \bibinfo {author} {\bibfnamefont {L.~F.}\ \bibnamefont {Buchmann}},
  \bibinfo {author} {\bibfnamefont {E.~M.}\ \bibnamefont {Wright}},\ and\
  \bibinfo {author} {\bibfnamefont {P.}~\bibnamefont {Meystre}},\ }\bibfield
  {title} {\bibinfo {title} {{``Multimode strong-coupling quantum
  optomechanics''}},\ }\href {https://doi.org/10.1103/PhysRevA.88.063850}
  {\bibfield  {journal} {\bibinfo  {journal} {Phys. Rev. A}\ }\textbf {\bibinfo
  {volume} {88}},\ \bibinfo {pages} {063850} (\bibinfo {year}
  {2013})}\BibitemShut {NoStop}%
\bibitem [{\citenamefont {Shkarin}\ \emph {et~al.}(2014)\citenamefont
  {Shkarin}, \citenamefont {Flowers-Jacobs}, \citenamefont {Hoch},
  \citenamefont {Kashkanova}, \citenamefont {Deutsch}, \citenamefont
  {Reichel},\ and\ \citenamefont {Harris}}]{Shkarin2014}%
  \BibitemOpen
  \bibfield  {author} {\bibinfo {author} {\bibfnamefont {A.~B.}\ \bibnamefont
  {Shkarin}}, \bibinfo {author} {\bibfnamefont {N.~E.}\ \bibnamefont
  {Flowers-Jacobs}}, \bibinfo {author} {\bibfnamefont {S.~W.}\ \bibnamefont
  {Hoch}}, \bibinfo {author} {\bibfnamefont {A.~D.}\ \bibnamefont
  {Kashkanova}}, \bibinfo {author} {\bibfnamefont {C.}~\bibnamefont {Deutsch}},
  \bibinfo {author} {\bibfnamefont {J.}~\bibnamefont {Reichel}},\ and\ \bibinfo
  {author} {\bibfnamefont {J.~G.~E.}\ \bibnamefont {Harris}},\ }\bibfield
  {title} {\bibinfo {title} {{``Optically Mediated Hybridization between Two
  Mechanical Modes''}},\ }\href
  {https://doi.org/10.1103/PhysRevLett.112.013602} {\bibfield  {journal}
  {\bibinfo  {journal} {Phys. Rev. Lett.}\ }\textbf {\bibinfo {volume} {112}},\
  \bibinfo {pages} {013602} (\bibinfo {year} {2014})}\BibitemShut {NoStop}%
\bibitem [{\citenamefont {Fan}\ \emph {et~al.}(2015)\citenamefont {Fan},
  \citenamefont {Fong}, \citenamefont {Poot},\ and\ \citenamefont
  {Tang}}]{Fan2015}%
  \BibitemOpen
  \bibfield  {author} {\bibinfo {author} {\bibfnamefont {L.}~\bibnamefont
  {Fan}}, \bibinfo {author} {\bibfnamefont {K.~Y.}\ \bibnamefont {Fong}},
  \bibinfo {author} {\bibfnamefont {M.}~\bibnamefont {Poot}},\ and\ \bibinfo
  {author} {\bibfnamefont {H.~X.}\ \bibnamefont {Tang}},\ }\bibfield  {title}
  {\bibinfo {title} {{``Cascaded optical transparency in multimode-cavity
  optomechanical systems''}},\ }\href {https://doi.org/10.1038/ncomms6850}
  {\bibfield  {journal} {\bibinfo  {journal} {Nat. Commun.}\ }\textbf {\bibinfo
  {volume} {6}},\ \bibinfo {pages} {5850} (\bibinfo {year} {2015})}\BibitemShut
  {NoStop}%
\bibitem [{\citenamefont {Nielsen}\ \emph {et~al.}(2017)\citenamefont
  {Nielsen}, \citenamefont {Tsaturyan}, \citenamefont {M{\o}ller},
  \citenamefont {Polzik},\ and\ \citenamefont {Schliesser}}]{Nielsen2017}%
  \BibitemOpen
  \bibfield  {author} {\bibinfo {author} {\bibfnamefont {W.~H.~P.}\
  \bibnamefont {Nielsen}}, \bibinfo {author} {\bibfnamefont {Y.}~\bibnamefont
  {Tsaturyan}}, \bibinfo {author} {\bibfnamefont {C.~B.}\ \bibnamefont
  {M{\o}ller}}, \bibinfo {author} {\bibfnamefont {E.~S.}\ \bibnamefont
  {Polzik}},\ and\ \bibinfo {author} {\bibfnamefont {A.}~\bibnamefont
  {Schliesser}},\ }\bibfield  {title} {\bibinfo {title} {{``Multimode
  optomechanical system in the quantum regime''}},\ }\href
  {https://doi.org/10.1073/pnas.1608412114} {\bibfield  {journal} {\bibinfo
  {journal} {Proc. Natl. Acad. Sci.}\ }\textbf {\bibinfo {volume} {114}},\
  \bibinfo {pages} {62} (\bibinfo {year} {2017})}\BibitemShut {NoStop}%
\bibitem [{\citenamefont {Arrangoiz-Arriola}\ \emph {et~al.}(2019)\citenamefont
  {Arrangoiz-Arriola}, \citenamefont {Wollack}, \citenamefont {Wang},
  \citenamefont {Pechal}, \citenamefont {Jiang}, \citenamefont {McKenna},
  \citenamefont {Witmer}, \citenamefont {Laer},\ and\ \citenamefont
  {Safavi-Naeini}}]{Arrangoiz2019}%
  \BibitemOpen
  \bibfield  {author} {\bibinfo {author} {\bibfnamefont {P.}~\bibnamefont
  {Arrangoiz-Arriola}}, \bibinfo {author} {\bibfnamefont {E.~A.}\ \bibnamefont
  {Wollack}}, \bibinfo {author} {\bibfnamefont {Z.}~\bibnamefont {Wang}},
  \bibinfo {author} {\bibfnamefont {M.}~\bibnamefont {Pechal}}, \bibinfo
  {author} {\bibfnamefont {W.}~\bibnamefont {Jiang}}, \bibinfo {author}
  {\bibfnamefont {T.~P.}\ \bibnamefont {McKenna}}, \bibinfo {author}
  {\bibfnamefont {J.~D.}\ \bibnamefont {Witmer}}, \bibinfo {author}
  {\bibfnamefont {R.~V.}\ \bibnamefont {Laer}},\ and\ \bibinfo {author}
  {\bibfnamefont {A.~H.}\ \bibnamefont {Safavi-Naeini}},\ }\bibfield  {title}
  {\bibinfo {title} {{``Resolving the energy levels of a nanomechanical
  oscillator''}},\ }\href {https://doi.org/10.1038/s41586-019-1386-x}
  {\bibfield  {journal} {\bibinfo  {journal} {Nature}\ }\textbf {\bibinfo
  {volume} {571}},\ \bibinfo {pages} {537} (\bibinfo {year}
  {2019})}\BibitemShut {NoStop}%
\bibitem [{\citenamefont {Han}\ \emph {et~al.}(2016)\citenamefont {Han},
  \citenamefont {Zou},\ and\ \citenamefont {Tang}}]{Han2016}%
  \BibitemOpen
  \bibfield  {author} {\bibinfo {author} {\bibfnamefont {X.}~\bibnamefont
  {Han}}, \bibinfo {author} {\bibfnamefont {C.-L.}\ \bibnamefont {Zou}},\ and\
  \bibinfo {author} {\bibfnamefont {H.~X.}\ \bibnamefont {Tang}},\ }\bibfield
  {title} {\bibinfo {title} {{``Multimode Strong Coupling in Superconducting
  Cavity Piezoelectromechanics''}},\ }\href
  {https://doi.org/10.1103/PhysRevLett.117.123603} {\bibfield  {journal}
  {\bibinfo  {journal} {Phys. Rev. Lett.}\ }\textbf {\bibinfo {volume} {117}},\
  \bibinfo {pages} {123603} (\bibinfo {year} {2016})}\BibitemShut {NoStop}%
\bibitem [{\citenamefont {Kervinen}\ \emph {et~al.}(2019)\citenamefont
  {Kervinen}, \citenamefont {Ramirez-Munoz}, \citenamefont {Valimaa},\ and\
  \citenamefont {Sillanpaa}}]{Kervinen2019}%
  \BibitemOpen
  \bibfield  {author} {\bibinfo {author} {\bibfnamefont {M.}~\bibnamefont
  {Kervinen}}, \bibinfo {author} {\bibfnamefont {J.~E.}\ \bibnamefont
  {Ramirez-Munoz}}, \bibinfo {author} {\bibfnamefont {A.}~\bibnamefont
  {Valimaa}},\ and\ \bibinfo {author} {\bibfnamefont {M.~A.}\ \bibnamefont
  {Sillanpaa}},\ }\bibfield  {title} {\bibinfo {title}
  {{``Landau-Zener-St{\"u}ckelberg Interference in a Multimode
  Electromechanical System in the Quantum Regime''}},\ }\href
  {https://doi.org/10.1103/physrevlett.123.240401} {\bibfield  {journal}
  {\bibinfo  {journal} {Phys. Rev. Lett.}\ }\textbf {\bibinfo {volume} {123}},\
  \bibinfo {pages} {240401} (\bibinfo {year} {2019})}\BibitemShut {NoStop}%
\bibitem [{\citenamefont {Gokhale}\ \emph {et~al.}(2020)\citenamefont
  {Gokhale}, \citenamefont {Downey}, \citenamefont {Katzer}, \citenamefont
  {Nepal}, \citenamefont {Lang}, \citenamefont {Stroud},\ and\ \citenamefont
  {Meyer}}]{Gokhale2020}%
  \BibitemOpen
  \bibfield  {author} {\bibinfo {author} {\bibfnamefont {V.~J.}\ \bibnamefont
  {Gokhale}}, \bibinfo {author} {\bibfnamefont {B.~P.}\ \bibnamefont {Downey}},
  \bibinfo {author} {\bibfnamefont {D.~S.}\ \bibnamefont {Katzer}}, \bibinfo
  {author} {\bibfnamefont {N.}~\bibnamefont {Nepal}}, \bibinfo {author}
  {\bibfnamefont {A.~C.}\ \bibnamefont {Lang}}, \bibinfo {author}
  {\bibfnamefont {R.~M.}\ \bibnamefont {Stroud}},\ and\ \bibinfo {author}
  {\bibfnamefont {D.~J.}\ \bibnamefont {Meyer}},\ }\bibfield  {title} {\bibinfo
  {title} {{``Epitaxial bulk acoustic wave resonators as highly coherent
  multi-phonon sources for quantum acoustodynamics''}},\ }\href
  {https://doi.org/10.1038/s41467-020-15472-w} {\bibfield  {journal} {\bibinfo
  {journal} {Nat. Commun.}\ }\textbf {\bibinfo {volume} {11}},\ \bibinfo
  {pages} {2314} (\bibinfo {year} {2020})}\BibitemShut {NoStop}%
\bibitem [{\citenamefont {Manenti}\ \emph {et~al.}(2017)\citenamefont
  {Manenti}, \citenamefont {Kockum}, \citenamefont {Patterson}, \citenamefont
  {Behrle}, \citenamefont {Rahamim}, \citenamefont {Tancredi}, \citenamefont
  {Nori},\ and\ \citenamefont {Leek}}]{Manenti2017}%
  \BibitemOpen
  \bibfield  {author} {\bibinfo {author} {\bibfnamefont {R.}~\bibnamefont
  {Manenti}}, \bibinfo {author} {\bibfnamefont {A.~F.}\ \bibnamefont {Kockum}},
  \bibinfo {author} {\bibfnamefont {A.}~\bibnamefont {Patterson}}, \bibinfo
  {author} {\bibfnamefont {T.}~\bibnamefont {Behrle}}, \bibinfo {author}
  {\bibfnamefont {J.}~\bibnamefont {Rahamim}}, \bibinfo {author} {\bibfnamefont
  {G.}~\bibnamefont {Tancredi}}, \bibinfo {author} {\bibfnamefont
  {F.}~\bibnamefont {Nori}},\ and\ \bibinfo {author} {\bibfnamefont {P.~J.}\
  \bibnamefont {Leek}},\ }\bibfield  {title} {\bibinfo {title} {{``Circuit
  quantum acoustodynamics with surface acoustic waves''}},\ }\href
  {https://doi.org/10.1038/s41467-017-01063-9} {\bibfield  {journal} {\bibinfo
  {journal} {Nat. Commun.}\ }\textbf {\bibinfo {volume} {8}},\ \bibinfo {pages}
  {975} (\bibinfo {year} {2017})}\BibitemShut {NoStop}%
\bibitem [{\citenamefont {Moores}\ \emph {et~al.}(2018)\citenamefont {Moores},
  \citenamefont {Sletten}, \citenamefont {Viennot},\ and\ \citenamefont
  {Lehnert}}]{Moores2018}%
  \BibitemOpen
  \bibfield  {author} {\bibinfo {author} {\bibfnamefont {B.~A.}\ \bibnamefont
  {Moores}}, \bibinfo {author} {\bibfnamefont {L.~R.}\ \bibnamefont {Sletten}},
  \bibinfo {author} {\bibfnamefont {J.~J.}\ \bibnamefont {Viennot}},\ and\
  \bibinfo {author} {\bibfnamefont {K.~W.}\ \bibnamefont {Lehnert}},\
  }\bibfield  {title} {\bibinfo {title} {{``Cavity Quantum Acoustic Device in
  the Multimode Strong Coupling Regime''}},\ }\href
  {https://doi.org/10.1103/physrevlett.120.227701} {\bibfield  {journal}
  {\bibinfo  {journal} {Phys. Rev. Lett.}\ }\textbf {\bibinfo {volume} {120}},\
  \bibinfo {pages} {227701} (\bibinfo {year} {2018})}\BibitemShut {NoStop}%
\bibitem [{\citenamefont {Satzinger}\ \emph {et~al.}(2018)\citenamefont
  {Satzinger}, \citenamefont {Zhong}, \citenamefont {Chang}, \citenamefont
  {Peairs}, \citenamefont {Bienfait}, \citenamefont {Chou}, \citenamefont
  {Cleland}, \citenamefont {Conner}, \citenamefont {Dumur}, \citenamefont
  {Grebel}, \citenamefont {Gutierrez}, \citenamefont {November}, \citenamefont
  {Povey}, \citenamefont {Whiteley}, \citenamefont {Awschalom}, \citenamefont
  {Schuster},\ and\ \citenamefont {Cleland}}]{Satzinger2018}%
  \BibitemOpen
  \bibfield  {author} {\bibinfo {author} {\bibfnamefont {K.~J.}\ \bibnamefont
  {Satzinger}}, \bibinfo {author} {\bibfnamefont {Y.~P.}\ \bibnamefont
  {Zhong}}, \bibinfo {author} {\bibfnamefont {H.-S.}\ \bibnamefont {Chang}},
  \bibinfo {author} {\bibfnamefont {G.~A.}\ \bibnamefont {Peairs}}, \bibinfo
  {author} {\bibfnamefont {A.}~\bibnamefont {Bienfait}}, \bibinfo {author}
  {\bibfnamefont {M.-H.}\ \bibnamefont {Chou}}, \bibinfo {author}
  {\bibfnamefont {A.~Y.}\ \bibnamefont {Cleland}}, \bibinfo {author}
  {\bibfnamefont {C.~R.}\ \bibnamefont {Conner}}, \bibinfo {author}
  {\bibfnamefont {{\'{E}}.}~\bibnamefont {Dumur}}, \bibinfo {author}
  {\bibfnamefont {J.}~\bibnamefont {Grebel}}, \bibinfo {author} {\bibfnamefont
  {I.}~\bibnamefont {Gutierrez}}, \bibinfo {author} {\bibfnamefont {B.~H.}\
  \bibnamefont {November}}, \bibinfo {author} {\bibfnamefont {R.~G.}\
  \bibnamefont {Povey}}, \bibinfo {author} {\bibfnamefont {S.~J.}\ \bibnamefont
  {Whiteley}}, \bibinfo {author} {\bibfnamefont {D.~D.}\ \bibnamefont
  {Awschalom}}, \bibinfo {author} {\bibfnamefont {D.~I.}\ \bibnamefont
  {Schuster}},\ and\ \bibinfo {author} {\bibfnamefont {A.~N.}\ \bibnamefont
  {Cleland}},\ }\bibfield  {title} {\bibinfo {title} {{``Quantum control of
  surface acoustic-wave phonons''}},\ }\href
  {https://doi.org/10.1038/s41586-018-0719-5} {\bibfield  {journal} {\bibinfo
  {journal} {Nature}\ }\textbf {\bibinfo {volume} {563}},\ \bibinfo {pages}
  {661} (\bibinfo {year} {2018})}\BibitemShut {NoStop}%
\bibitem [{\citenamefont {Sletten}\ \emph {et~al.}(2019)\citenamefont
  {Sletten}, \citenamefont {Moores}, \citenamefont {Viennot},\ and\
  \citenamefont {Lehnert}}]{Sletten2019}%
  \BibitemOpen
  \bibfield  {author} {\bibinfo {author} {\bibfnamefont {L.~R.}\ \bibnamefont
  {Sletten}}, \bibinfo {author} {\bibfnamefont {B.~A.}\ \bibnamefont {Moores}},
  \bibinfo {author} {\bibfnamefont {J.~J.}\ \bibnamefont {Viennot}},\ and\
  \bibinfo {author} {\bibfnamefont {K.~W.}\ \bibnamefont {Lehnert}},\
  }\bibfield  {title} {\bibinfo {title} {{``Resolving Phonon Fock States in a
  Multimode Cavity with a Double-Slit Qubit''}},\ }\href
  {https://doi.org/10.1103/physrevx.9.021056} {\bibfield  {journal} {\bibinfo
  {journal} {Phys. Rev. X}\ }\textbf {\bibinfo {volume} {9}},\ \bibinfo {pages}
  {021056} (\bibinfo {year} {2019})}\BibitemShut {NoStop}%
\bibitem [{\citenamefont {Bienfait}\ \emph {et~al.}(2019)\citenamefont
  {Bienfait}, \citenamefont {Satzinger}, \citenamefont {Zhong}, \citenamefont
  {Chang}, \citenamefont {Chou}, \citenamefont {Conner}, \citenamefont {Dumur},
  \citenamefont {Grebel}, \citenamefont {Peairs}, \citenamefont {Povey},\ and\
  \citenamefont {Cleland}}]{Bienfait2019}%
  \BibitemOpen
  \bibfield  {author} {\bibinfo {author} {\bibfnamefont {A.}~\bibnamefont
  {Bienfait}}, \bibinfo {author} {\bibfnamefont {K.~J.}\ \bibnamefont
  {Satzinger}}, \bibinfo {author} {\bibfnamefont {Y.~P.}\ \bibnamefont
  {Zhong}}, \bibinfo {author} {\bibfnamefont {H.-S.}\ \bibnamefont {Chang}},
  \bibinfo {author} {\bibfnamefont {M.-H.}\ \bibnamefont {Chou}}, \bibinfo
  {author} {\bibfnamefont {C.~R.}\ \bibnamefont {Conner}}, \bibinfo {author}
  {\bibfnamefont {{\'{E}}.}~\bibnamefont {Dumur}}, \bibinfo {author}
  {\bibfnamefont {J.}~\bibnamefont {Grebel}}, \bibinfo {author} {\bibfnamefont
  {G.~A.}\ \bibnamefont {Peairs}}, \bibinfo {author} {\bibfnamefont {R.~G.}\
  \bibnamefont {Povey}},\ and\ \bibinfo {author} {\bibfnamefont {A.~N.}\
  \bibnamefont {Cleland}},\ }\bibfield  {title} {\bibinfo {title}
  {{``Phonon-mediated quantum state transfer and remote qubit
  entanglement''}},\ }\href {https://doi.org/10.1126/science.aaw8415}
  {\bibfield  {journal} {\bibinfo  {journal} {Science}\ }\textbf {\bibinfo
  {volume} {364}},\ \bibinfo {pages} {368} (\bibinfo {year}
  {2019})}\BibitemShut {NoStop}%
\bibitem [{\citenamefont {Houhou}\ \emph {et~al.}(2015)\citenamefont {Houhou},
  \citenamefont {Aissaoui},\ and\ \citenamefont {Ferraro}}]{Houhou2015}%
  \BibitemOpen
  \bibfield  {author} {\bibinfo {author} {\bibfnamefont {O.}~\bibnamefont
  {Houhou}}, \bibinfo {author} {\bibfnamefont {H.}~\bibnamefont {Aissaoui}},\
  and\ \bibinfo {author} {\bibfnamefont {A.}~\bibnamefont {Ferraro}},\
  }\bibfield  {title} {\bibinfo {title} {{``Generation of cluster states in
  optomechanical quantum systems''}},\ }\href
  {https://doi.org/10.1103/PhysRevA.92.063843} {\bibfield  {journal} {\bibinfo
  {journal} {Phys. Rev. A}\ }\textbf {\bibinfo {volume} {92}},\ \bibinfo
  {pages} {063843} (\bibinfo {year} {2015})}\BibitemShut {NoStop}%
\bibitem [{\citenamefont {Moore}\ \emph {et~al.}(2017)\citenamefont {Moore},
  \citenamefont {Houhou},\ and\ \citenamefont {Ferraro}}]{Moore2017}%
  \BibitemOpen
  \bibfield  {author} {\bibinfo {author} {\bibfnamefont {D.~W.}\ \bibnamefont
  {Moore}}, \bibinfo {author} {\bibfnamefont {O.}~\bibnamefont {Houhou}},\ and\
  \bibinfo {author} {\bibfnamefont {A.}~\bibnamefont {Ferraro}},\ }\bibfield
  {title} {\bibinfo {title} {{``Arbitrary multimode Gaussian operations on
  mechanical cluster states''}},\ }\href
  {https://doi.org/10.1103/PhysRevA.96.022305} {\bibfield  {journal} {\bibinfo
  {journal} {Phys. Rev. A}\ }\textbf {\bibinfo {volume} {96}},\ \bibinfo
  {pages} {022305} (\bibinfo {year} {2017})}\BibitemShut {NoStop}%
\bibitem [{\citenamefont {Tan}\ \emph {et~al.}(2017)\citenamefont {Tan},
  \citenamefont {Wei},\ and\ \citenamefont {Li}}]{Tan2017}%
  \BibitemOpen
  \bibfield  {author} {\bibinfo {author} {\bibfnamefont {H.}~\bibnamefont
  {Tan}}, \bibinfo {author} {\bibfnamefont {Y.}~\bibnamefont {Wei}},\ and\
  \bibinfo {author} {\bibfnamefont {G.}~\bibnamefont {Li}},\ }\bibfield
  {title} {\bibinfo {title} {{``Building mechanical Greenberger-Horne-Zeilinger
  and cluster states by harnessing optomechanical quantum steerable
  correlations''}},\ }\href {https://doi.org/10.1103/PhysRevA.96.052331}
  {\bibfield  {journal} {\bibinfo  {journal} {Phys. Rev. A}\ }\textbf {\bibinfo
  {volume} {96}},\ \bibinfo {pages} {052331} (\bibinfo {year}
  {2017})}\BibitemShut {NoStop}%
\bibitem [{\citenamefont {Houhou}\ \emph {et~al.}()\citenamefont {Houhou},
  \citenamefont {Moore}, \citenamefont {Bose},\ and\ \citenamefont
  {Ferraro}}]{Houhou2018}%
  \BibitemOpen
  \bibfield  {author} {\bibinfo {author} {\bibfnamefont {O.}~\bibnamefont
  {Houhou}}, \bibinfo {author} {\bibfnamefont {D.~W.}\ \bibnamefont {Moore}},
  \bibinfo {author} {\bibfnamefont {S.}~\bibnamefont {Bose}},\ and\ \bibinfo
  {author} {\bibfnamefont {A.}~\bibnamefont {Ferraro}},\ }\bibfield  {title}
  {\bibinfo {title} {{``Unconditional measurement-based quantum computation
  with optomechanical continuous variables''}},\ }\href
  {https://arxiv.org/abs/1809.09733} {\ }\Eprint
  {https://arxiv.org/abs/1809.09733} {arXiv:1809.09733 [quant-ph]} \BibitemShut
  {NoStop}%
\bibitem [{\citenamefont {Hann}\ \emph {et~al.}(2019)\citenamefont {Hann},
  \citenamefont {Zou}, \citenamefont {Zhang}, \citenamefont {Chu},
  \citenamefont {Schoelkopf}, \citenamefont {Girvin},\ and\ \citenamefont
  {Jiang}}]{Hann2019}%
  \BibitemOpen
  \bibfield  {author} {\bibinfo {author} {\bibfnamefont {C.~T.}\ \bibnamefont
  {Hann}}, \bibinfo {author} {\bibfnamefont {C.-L.}\ \bibnamefont {Zou}},
  \bibinfo {author} {\bibfnamefont {Y.}~\bibnamefont {Zhang}}, \bibinfo
  {author} {\bibfnamefont {Y.}~\bibnamefont {Chu}}, \bibinfo {author}
  {\bibfnamefont {R.~J.}\ \bibnamefont {Schoelkopf}}, \bibinfo {author}
  {\bibfnamefont {S.~M.}\ \bibnamefont {Girvin}},\ and\ \bibinfo {author}
  {\bibfnamefont {L.}~\bibnamefont {Jiang}},\ }\bibfield  {title} {\bibinfo
  {title} {{``Hardware-Efficient Quantum Random Access Memory with Hybrid
  Quantum Acoustic Systems''}},\ }\href
  {https://doi.org/10.1103/physrevlett.123.250501} {\bibfield  {journal}
  {\bibinfo  {journal} {Physical Review Letters}\ }\textbf {\bibinfo {volume}
  {123}},\ \bibinfo {pages} {250501} (\bibinfo {year} {2019})}\BibitemShut
  {NoStop}%
\bibitem [{\citenamefont {Chu}\ \emph {et~al.}(2017)\citenamefont {Chu},
  \citenamefont {Kharel}, \citenamefont {Renninger}, \citenamefont {Burkhart},
  \citenamefont {Frunzio}, \citenamefont {Rakich},\ and\ \citenamefont
  {Schoelkopf}}]{Chu2017}%
  \BibitemOpen
  \bibfield  {author} {\bibinfo {author} {\bibfnamefont {Y.}~\bibnamefont
  {Chu}}, \bibinfo {author} {\bibfnamefont {P.}~\bibnamefont {Kharel}},
  \bibinfo {author} {\bibfnamefont {W.~H.}\ \bibnamefont {Renninger}}, \bibinfo
  {author} {\bibfnamefont {L.~D.}\ \bibnamefont {Burkhart}}, \bibinfo {author}
  {\bibfnamefont {L.}~\bibnamefont {Frunzio}}, \bibinfo {author} {\bibfnamefont
  {P.~T.}\ \bibnamefont {Rakich}},\ and\ \bibinfo {author} {\bibfnamefont
  {R.~J.}\ \bibnamefont {Schoelkopf}},\ }\bibfield  {title} {\bibinfo {title}
  {{``Quantum acoustics with superconducting qubits''}},\ }\href
  {https://doi.org/10.1126/science.aao1511} {\bibfield  {journal} {\bibinfo
  {journal} {Science}\ }\textbf {\bibinfo {volume} {358}},\ \bibinfo {pages}
  {199} (\bibinfo {year} {2017})}\BibitemShut {NoStop}%
\bibitem [{\citenamefont {Andersson}\ \emph {et~al.}()\citenamefont
  {Andersson}, \citenamefont {Jolin}, \citenamefont {Scigliuzzo}, \citenamefont
  {Borgani}, \citenamefont {Thol{\'e}n}, \citenamefont {Haviland},\ and\
  \citenamefont {Delsing}}]{Andersson2020}%
  \BibitemOpen
  \bibfield  {author} {\bibinfo {author} {\bibfnamefont {G.}~\bibnamefont
  {Andersson}}, \bibinfo {author} {\bibfnamefont {S.~W.}\ \bibnamefont
  {Jolin}}, \bibinfo {author} {\bibfnamefont {M.}~\bibnamefont {Scigliuzzo}},
  \bibinfo {author} {\bibfnamefont {R.}~\bibnamefont {Borgani}}, \bibinfo
  {author} {\bibfnamefont {M.~O.}\ \bibnamefont {Thol{\'e}n}}, \bibinfo
  {author} {\bibfnamefont {D.~B.}\ \bibnamefont {Haviland}},\ and\ \bibinfo
  {author} {\bibfnamefont {P.}~\bibnamefont {Delsing}},\ }\bibfield  {title}
  {\bibinfo {title} {{``Squeezing and correlations of multiple modes in a
  parametric acoustic cavity''}},\ }\href@noop {} {\ }\Eprint
  {https://arxiv.org/abs/2007.05826} {arXiv:2007.05826} \BibitemShut {NoStop}%
\bibitem [{\citenamefont {Abdi}\ \emph {et~al.}(2018)\citenamefont {Abdi},
  \citenamefont {Chou}, \citenamefont {Gali},\ and\ \citenamefont
  {Plenio}}]{Abdi2018a}%
  \BibitemOpen
  \bibfield  {author} {\bibinfo {author} {\bibfnamefont {M.}~\bibnamefont
  {Abdi}}, \bibinfo {author} {\bibfnamefont {J.-P.}\ \bibnamefont {Chou}},
  \bibinfo {author} {\bibfnamefont {A.}~\bibnamefont {Gali}},\ and\ \bibinfo
  {author} {\bibfnamefont {M.~B.}\ \bibnamefont {Plenio}},\ }\bibfield  {title}
  {\bibinfo {title} {{``Color Centers in Hexagonal Boron Nitride Monolayers: A
  Group Theory and Ab Initio Analysis''}},\ }\href
  {https://doi.org/10.1021/acsphotonics.7b01442} {\bibfield  {journal}
  {\bibinfo  {journal} {{ACS} Photonics}\ }\textbf {\bibinfo {volume} {5}},\
  \bibinfo {pages} {1967} (\bibinfo {year} {2018})}\BibitemShut {NoStop}%
\bibitem [{\citenamefont {Abdi}\ and\ \citenamefont
  {Plenio}(2018)}]{Abdi2018b}%
  \BibitemOpen
  \bibfield  {author} {\bibinfo {author} {\bibfnamefont {M.}~\bibnamefont
  {Abdi}}\ and\ \bibinfo {author} {\bibfnamefont {M.~B.}\ \bibnamefont
  {Plenio}},\ }\bibfield  {title} {\bibinfo {title} {{``Analog quantum
  simulation of extremely sub-Ohmic spin-boson models''}},\ }\href
  {https://doi.org/10.1103/PhysRevA.98.040303} {\bibfield  {journal} {\bibinfo
  {journal} {Phys. Rev. A}\ }\textbf {\bibinfo {volume} {98}},\ \bibinfo
  {pages} {040303(R)} (\bibinfo {year} {2018})}\BibitemShut {NoStop}%
\bibitem [{\citenamefont {Goryachev}\ and\ \citenamefont
  {Tobar}(2014)}]{Goryachev2014}%
  \BibitemOpen
  \bibfield  {author} {\bibinfo {author} {\bibfnamefont {M.}~\bibnamefont
  {Goryachev}}\ and\ \bibinfo {author} {\bibfnamefont {M.~E.}\ \bibnamefont
  {Tobar}},\ }\bibfield  {title} {\bibinfo {title} {{``Gravitational wave
  detection with high frequency phonon trapping acoustic cavities''}},\ }\href
  {https://doi.org/10.1103/physrevd.90.102005} {\bibfield  {journal} {\bibinfo
  {journal} {Phys. Rev. D}\ }\textbf {\bibinfo {volume} {90}},\ \bibinfo
  {pages} {102005} (\bibinfo {year} {2014})}\BibitemShut {NoStop}%
\bibitem [{\citenamefont {Arvanitaki}\ \emph {et~al.}(2016)\citenamefont
  {Arvanitaki}, \citenamefont {Dimopoulos},\ and\ \citenamefont {{Van
  Tilburg}}}]{Arvanitaki2016}%
  \BibitemOpen
  \bibfield  {author} {\bibinfo {author} {\bibfnamefont {A.}~\bibnamefont
  {Arvanitaki}}, \bibinfo {author} {\bibfnamefont {S.}~\bibnamefont
  {Dimopoulos}},\ and\ \bibinfo {author} {\bibfnamefont {K.}~\bibnamefont {{Van
  Tilburg}}},\ }\bibfield  {title} {\bibinfo {title} {{``Sound of Dark Matter:
  Searching for Light Scalars with Resonant-Mass Detectors''}},\ }\href
  {https://doi.org/10.1103/physrevlett.116.031102} {\bibfield  {journal}
  {\bibinfo  {journal} {Phys. Rev. Lett.}\ }\textbf {\bibinfo {volume} {116}},\
  \bibinfo {pages} {031102} (\bibinfo {year} {2016})}\BibitemShut {NoStop}%
\bibitem [{\citenamefont {Abdi}\ \emph {et~al.}(2016)\citenamefont {Abdi},
  \citenamefont {Degenfeld-Schonburg}, \citenamefont {Sameti}, \citenamefont
  {Navarrete-Benlloch},\ and\ \citenamefont {Hartmann}}]{Abdi2016}%
  \BibitemOpen
  \bibfield  {author} {\bibinfo {author} {\bibfnamefont {M.}~\bibnamefont
  {Abdi}}, \bibinfo {author} {\bibfnamefont {P.}~\bibnamefont
  {Degenfeld-Schonburg}}, \bibinfo {author} {\bibfnamefont {M.}~\bibnamefont
  {Sameti}}, \bibinfo {author} {\bibfnamefont {C.}~\bibnamefont
  {Navarrete-Benlloch}},\ and\ \bibinfo {author} {\bibfnamefont {M.~J.}\
  \bibnamefont {Hartmann}},\ }\bibfield  {title} {\bibinfo {title}
  {{``Dissipative Optomechanical Preparation of Macroscopic Quantum
  Superposition States''}},\ }\href
  {https://doi.org/10.1103/PhysRevLett.116.233604} {\bibfield  {journal}
  {\bibinfo  {journal} {Phys. Rev. Lett.}\ }\textbf {\bibinfo {volume} {116}},\
  \bibinfo {pages} {233604} (\bibinfo {year} {2016})}\BibitemShut {NoStop}%
\bibitem [{\citenamefont {Renninger}\ \emph {et~al.}(2018)\citenamefont
  {Renninger}, \citenamefont {Kharel}, \citenamefont {Behunin},\ and\
  \citenamefont {Rakich}}]{Renninger2018}%
  \BibitemOpen
  \bibfield  {author} {\bibinfo {author} {\bibfnamefont {W.~H.}\ \bibnamefont
  {Renninger}}, \bibinfo {author} {\bibfnamefont {P.}~\bibnamefont {Kharel}},
  \bibinfo {author} {\bibfnamefont {R.~O.}\ \bibnamefont {Behunin}},\ and\
  \bibinfo {author} {\bibfnamefont {P.~T.}\ \bibnamefont {Rakich}},\ }\bibfield
   {title} {\bibinfo {title} {Bulk crystalline optomechanics},\ }\href
  {https://doi.org/10.1038/s41567-018-0090-3} {\bibfield  {journal} {\bibinfo
  {journal} {Nat. Phys.}\ }\textbf {\bibinfo {volume} {14}},\ \bibinfo {pages}
  {601} (\bibinfo {year} {2018})}\BibitemShut {NoStop}%
\bibitem [{\citenamefont {Kolkowitz}\ \emph {et~al.}(2012)\citenamefont
  {Kolkowitz}, \citenamefont {Jayich}, \citenamefont {Unterreithmeier},
  \citenamefont {Bennett}, \citenamefont {Rabl}, \citenamefont {Harris},\ and\
  \citenamefont {Lukin}}]{Kolkowitz2012}%
  \BibitemOpen
  \bibfield  {author} {\bibinfo {author} {\bibfnamefont {S.}~\bibnamefont
  {Kolkowitz}}, \bibinfo {author} {\bibfnamefont {A.~C.~B.}\ \bibnamefont
  {Jayich}}, \bibinfo {author} {\bibfnamefont {Q.~P.}\ \bibnamefont
  {Unterreithmeier}}, \bibinfo {author} {\bibfnamefont {S.~D.}\ \bibnamefont
  {Bennett}}, \bibinfo {author} {\bibfnamefont {P.}~\bibnamefont {Rabl}},
  \bibinfo {author} {\bibfnamefont {J.~G.~E.}\ \bibnamefont {Harris}},\ and\
  \bibinfo {author} {\bibfnamefont {M.~D.}\ \bibnamefont {Lukin}},\ }\bibfield
  {title} {\bibinfo {title} {{``Coherent Sensing of a Mechanical Resonator with
  a Single-Spin Qubit''}},\ }\href {https://doi.org/10.1126/science.1216821}
  {\bibfield  {journal} {\bibinfo  {journal} {Science}\ }\textbf {\bibinfo
  {volume} {335}},\ \bibinfo {pages} {1603} (\bibinfo {year}
  {2012})}\BibitemShut {NoStop}%
\bibitem [{\citenamefont {Muschik}\ \emph {et~al.}(2014)\citenamefont
  {Muschik}, \citenamefont {Moulieras}, \citenamefont {Bachtold}, \citenamefont
  {Koppens}, \citenamefont {Lewenstein},\ and\ \citenamefont
  {Chang}}]{Muschik2014}%
  \BibitemOpen
  \bibfield  {author} {\bibinfo {author} {\bibfnamefont {C.~A.}\ \bibnamefont
  {Muschik}}, \bibinfo {author} {\bibfnamefont {S.}~\bibnamefont {Moulieras}},
  \bibinfo {author} {\bibfnamefont {A.}~\bibnamefont {Bachtold}}, \bibinfo
  {author} {\bibfnamefont {F.~H.~L.}\ \bibnamefont {Koppens}}, \bibinfo
  {author} {\bibfnamefont {M.}~\bibnamefont {Lewenstein}},\ and\ \bibinfo
  {author} {\bibfnamefont {D.~E.}\ \bibnamefont {Chang}},\ }\bibfield  {title}
  {\bibinfo {title} {{``Harnessing Vacuum Forces for Quantum Sensing of
  Graphene Motion''}},\ }\href {https://doi.org/10.1103/PhysRevLett.112.223601}
  {\bibfield  {journal} {\bibinfo  {journal} {Phys. Rev. Lett.}\ }\textbf
  {\bibinfo {volume} {112}},\ \bibinfo {pages} {223601} (\bibinfo {year}
  {2014})}\BibitemShut {NoStop}%
\bibitem [{\citenamefont {Li}\ \emph {et~al.}(2016)\citenamefont {Li},
  \citenamefont {Xiang}, \citenamefont {Rabl},\ and\ \citenamefont
  {Nori}}]{Li2016}%
  \BibitemOpen
  \bibfield  {author} {\bibinfo {author} {\bibfnamefont {P.-B.}\ \bibnamefont
  {Li}}, \bibinfo {author} {\bibfnamefont {Z.-L.}\ \bibnamefont {Xiang}},
  \bibinfo {author} {\bibfnamefont {P.}~\bibnamefont {Rabl}},\ and\ \bibinfo
  {author} {\bibfnamefont {F.}~\bibnamefont {Nori}},\ }\bibfield  {title}
  {\bibinfo {title} {{``Hybrid Quantum Device with Nitrogen-Vacancy Centers in
  Diamond Coupled to Carbon Nanotubes''}},\ }\href
  {https://doi.org/10.1103/PhysRevLett.117.015502} {\bibfield  {journal}
  {\bibinfo  {journal} {Phys. Rev. Lett.}\ }\textbf {\bibinfo {volume} {117}},\
  \bibinfo {pages} {015502} (\bibinfo {year} {2016})}\BibitemShut {NoStop}%
\bibitem [{\citenamefont {Abdi}\ \emph {et~al.}(2017)\citenamefont {Abdi},
  \citenamefont {Hwang}, \citenamefont {Aghtar},\ and\ \citenamefont
  {Plenio}}]{Abdi2017}%
  \BibitemOpen
  \bibfield  {author} {\bibinfo {author} {\bibfnamefont {M.}~\bibnamefont
  {Abdi}}, \bibinfo {author} {\bibfnamefont {M.-J.}\ \bibnamefont {Hwang}},
  \bibinfo {author} {\bibfnamefont {M.}~\bibnamefont {Aghtar}},\ and\ \bibinfo
  {author} {\bibfnamefont {M.~B.}\ \bibnamefont {Plenio}},\ }\bibfield  {title}
  {\bibinfo {title} {{``Spin-Mechanical Scheme with Color Centers in Hexagonal
  Boron Nitride Membranes''}},\ }\href
  {https://doi.org/10.1103/PhysRevLett.119.233602} {\bibfield  {journal}
  {\bibinfo  {journal} {Phys. Rev. Lett.}\ }\textbf {\bibinfo {volume} {119}},\
  \bibinfo {pages} {233602} (\bibinfo {year} {2017})}\BibitemShut {NoStop}%
\bibitem [{\citenamefont {Abdi}\ and\ \citenamefont {Plenio}(2019)}]{Abdi2019}%
  \BibitemOpen
  \bibfield  {author} {\bibinfo {author} {\bibfnamefont {M.}~\bibnamefont
  {Abdi}}\ and\ \bibinfo {author} {\bibfnamefont {M.~B.}\ \bibnamefont
  {Plenio}},\ }\bibfield  {title} {\bibinfo {title} {{``Quantum Effects in a
  Mechanically Modulated Single-Photon Emitter''}},\ }\href
  {https://doi.org/10.1103/PhysRevLett.122.023602} {\bibfield  {journal}
  {\bibinfo  {journal} {Phys. Rev. Lett.}\ }\textbf {\bibinfo {volume} {122}},\
  \bibinfo {pages} {023602} (\bibinfo {year} {2019})}\BibitemShut {NoStop}%
\bibitem [{\citenamefont {Rabl}(2010)}]{Rabl2010a}%
  \BibitemOpen
  \bibfield  {author} {\bibinfo {author} {\bibfnamefont {P.}~\bibnamefont
  {Rabl}},\ }\bibfield  {title} {\bibinfo {title} {{``Cooling of mechanical
  motion with a two-level system: The high-temperature regime''}},\ }\href
  {https://doi.org/10.1103/PhysRevB.82.165320} {\bibfield  {journal} {\bibinfo
  {journal} {Phys. Rev. B}\ }\textbf {\bibinfo {volume} {82}},\ \bibinfo
  {pages} {165320} (\bibinfo {year} {2010})}\BibitemShut {NoStop}%
\bibitem [{\citenamefont {Galve}\ \emph {et~al.}(2010)\citenamefont {Galve},
  \citenamefont {Pach{\'o}n},\ and\ \citenamefont {Zueco}}]{Galve2010}%
  \BibitemOpen
  \bibfield  {author} {\bibinfo {author} {\bibfnamefont {F.}~\bibnamefont
  {Galve}}, \bibinfo {author} {\bibfnamefont {L.~A.}\ \bibnamefont
  {Pach{\'o}n}},\ and\ \bibinfo {author} {\bibfnamefont {D.}~\bibnamefont
  {Zueco}},\ }\bibfield  {title} {\bibinfo {title} {{``Bringing Entanglement to
  the High Temperature Limit''}},\ }\href
  {https://doi.org/PhysRevLett.105.180501} {\bibfield  {journal} {\bibinfo
  {journal} {Phys. Rev. Lett.}\ }\textbf {\bibinfo {volume} {105}},\ \bibinfo
  {pages} {180501} (\bibinfo {year} {2010})}\BibitemShut {NoStop}%
\bibitem [{\citenamefont {Pfister}\ \emph {et~al.}(2004)\citenamefont
  {Pfister}, \citenamefont {Feng}, \citenamefont {Jennings}, \citenamefont
  {Pooser},\ and\ \citenamefont {Xie}}]{Pfister2004}%
  \BibitemOpen
  \bibfield  {author} {\bibinfo {author} {\bibfnamefont {O.}~\bibnamefont
  {Pfister}}, \bibinfo {author} {\bibfnamefont {S.}~\bibnamefont {Feng}},
  \bibinfo {author} {\bibfnamefont {G.}~\bibnamefont {Jennings}}, \bibinfo
  {author} {\bibfnamefont {R.}~\bibnamefont {Pooser}},\ and\ \bibinfo {author}
  {\bibfnamefont {D.}~\bibnamefont {Xie}},\ }\bibfield  {title} {\bibinfo
  {title} {{``Multipartite continuous-variable entanglement from concurrent
  nonlinearities''}},\ }\href {https://doi.org/10.1103/PhysRevA.70.020302}
  {\bibfield  {journal} {\bibinfo  {journal} {Phys. Rev. A}\ }\textbf {\bibinfo
  {volume} {70}},\ \bibinfo {pages} {020302(R)} (\bibinfo {year}
  {2004})}\BibitemShut {NoStop}%
\bibitem [{\citenamefont {Bradley}\ \emph {et~al.}(2005)\citenamefont
  {Bradley}, \citenamefont {Olsen}, \citenamefont {Pfister},\ and\
  \citenamefont {Pooser}}]{Bradley2005}%
  \BibitemOpen
  \bibfield  {author} {\bibinfo {author} {\bibfnamefont {A.~S.}\ \bibnamefont
  {Bradley}}, \bibinfo {author} {\bibfnamefont {M.~K.}\ \bibnamefont {Olsen}},
  \bibinfo {author} {\bibfnamefont {O.}~\bibnamefont {Pfister}},\ and\ \bibinfo
  {author} {\bibfnamefont {R.~C.}\ \bibnamefont {Pooser}},\ }\bibfield  {title}
  {\bibinfo {title} {{``Bright tripartite entanglement in triply concurrent
  parametric oscillation''}},\ }\href
  {https://doi.org/10.1103/PhysRevA.72.053805} {\bibfield  {journal} {\bibinfo
  {journal} {Phys. Rev. A}\ }\textbf {\bibinfo {volume} {72}},\ \bibinfo
  {pages} {053805} (\bibinfo {year} {2005})}\BibitemShut {NoStop}%
\bibitem [{\citenamefont {Zhang}\ and\ \citenamefont
  {Braunstein}(2006)}]{Zhang2006}%
  \BibitemOpen
  \bibfield  {author} {\bibinfo {author} {\bibfnamefont {J.}~\bibnamefont
  {Zhang}}\ and\ \bibinfo {author} {\bibfnamefont {S.~L.}\ \bibnamefont
  {Braunstein}},\ }\bibfield  {title} {\bibinfo {title} {{``Continuous-variable
  Gaussian analog of cluster states''}},\ }\href
  {https://doi.org/10.1103/PhysRevA.73.032318} {\bibfield  {journal} {\bibinfo
  {journal} {Phys. Rev. A}\ }\textbf {\bibinfo {volume} {73}},\ \bibinfo
  {pages} {032318} (\bibinfo {year} {2006})}\BibitemShut {NoStop}%
\bibitem [{\citenamefont {Briegel}\ \emph {et~al.}(2009)\citenamefont
  {Briegel}, \citenamefont {Browne}, \citenamefont {D{\"u}r}, \citenamefont
  {Raussendorf},\ and\ \citenamefont {{Van den Nest}}}]{Briegel2009}%
  \BibitemOpen
  \bibfield  {author} {\bibinfo {author} {\bibfnamefont {H.~J.}\ \bibnamefont
  {Briegel}}, \bibinfo {author} {\bibfnamefont {D.~E.}\ \bibnamefont {Browne}},
  \bibinfo {author} {\bibfnamefont {W.}~\bibnamefont {D{\"u}r}}, \bibinfo
  {author} {\bibfnamefont {R.}~\bibnamefont {Raussendorf}},\ and\ \bibinfo
  {author} {\bibfnamefont {M.}~\bibnamefont {{Van den Nest}}},\ }\bibfield
  {title} {\bibinfo {title} {{``Measurement-based quantum computation''}},\
  }\href {https://doi.org/10.1038/NPHYS1157} {\bibfield  {journal} {\bibinfo
  {journal} {Nat. Phys.}\ }\textbf {\bibinfo {volume} {5}},\ \bibinfo {pages}
  {19} (\bibinfo {year} {2009})}\BibitemShut {NoStop}%
\bibitem [{\citenamefont {Kervinen}\ \emph {et~al.}(2020)\citenamefont
  {Kervinen}, \citenamefont {Valimaa}, \citenamefont {Ramirez-Munoz},\ and\
  \citenamefont {Sillanpaa}}]{Kervinen2020}%
  \BibitemOpen
  \bibfield  {author} {\bibinfo {author} {\bibfnamefont {M.}~\bibnamefont
  {Kervinen}}, \bibinfo {author} {\bibfnamefont {A.}~\bibnamefont {Valimaa}},
  \bibinfo {author} {\bibfnamefont {J.~E.}\ \bibnamefont {Ramirez-Munoz}},\
  and\ \bibinfo {author} {\bibfnamefont {M.~A.}\ \bibnamefont {Sillanpaa}},\
  }\bibfield  {title} {\bibinfo {title} {{``Sideband Control of a Multimode
  Quantum Bulk Acoustic System''}},\ }\href
  {https://doi.org/10.1103/physrevapplied.14.054023} {\bibfield  {journal}
  {\bibinfo  {journal} {Phys. Rev. Appl}\ }\textbf {\bibinfo {volume} {14}},\
  \bibinfo {pages} {054023} (\bibinfo {year} {2020})}\BibitemShut {NoStop}%
\bibitem [{\citenamefont {Adesso}\ and\ \citenamefont
  {Illuminati}(2008)}]{Adesso2008}%
  \BibitemOpen
  \bibfield  {author} {\bibinfo {author} {\bibfnamefont {G.}~\bibnamefont
  {Adesso}}\ and\ \bibinfo {author} {\bibfnamefont {F.}~\bibnamefont
  {Illuminati}},\ }\bibfield  {title} {\bibinfo {title} {{``Genuine
  multipartite entanglement of symmetric Gaussian states: Strong monogamy,
  unitary localization, scaling behavior, and molecular sharing structure''}},\
  }\href {https://doi.org/10.1103/PhysRevA.78.042310} {\bibfield  {journal}
  {\bibinfo  {journal} {Phys. Rev. A}\ }\textbf {\bibinfo {volume} {78}},\
  \bibinfo {pages} {042310} (\bibinfo {year} {2008})}\BibitemShut {NoStop}%
\bibitem [{\citenamefont {Braunstein}\ and\ \citenamefont
  {Caves}(1994)}]{Braunstein1994}%
  \BibitemOpen
  \bibfield  {author} {\bibinfo {author} {\bibfnamefont {S.~L.}\ \bibnamefont
  {Braunstein}}\ and\ \bibinfo {author} {\bibfnamefont {C.~M.}\ \bibnamefont
  {Caves}},\ }\bibfield  {title} {\bibinfo {title} {{``Statistical Distance and
  the Geometry of Quantum States''}},\ }\href
  {https://doi.org/10.1103/PhysRevLett.72.3439} {\bibfield  {journal} {\bibinfo
   {journal} {Phys. Rev. Lett.}\ }\textbf {\bibinfo {volume} {72}},\ \bibinfo
  {pages} {3439} (\bibinfo {year} {1994})}\BibitemShut {NoStop}%
\bibitem [{\citenamefont {Genes}\ \emph {et~al.}(2008)\citenamefont {Genes},
  \citenamefont {Mari}, \citenamefont {Tombesi},\ and\ \citenamefont
  {Vitali}}]{Genes2008}%
  \BibitemOpen
  \bibfield  {author} {\bibinfo {author} {\bibfnamefont {C.}~\bibnamefont
  {Genes}}, \bibinfo {author} {\bibfnamefont {A.}~\bibnamefont {Mari}},
  \bibinfo {author} {\bibfnamefont {P.}~\bibnamefont {Tombesi}},\ and\ \bibinfo
  {author} {\bibfnamefont {D.}~\bibnamefont {Vitali}},\ }\bibfield  {title}
  {\bibinfo {title} {{``Robust entanglement of a micromechanical resonator with
  output optical fields''}},\ }\href
  {https://doi.org/10.1103/PhysRevA.78.032316} {\bibfield  {journal} {\bibinfo
  {journal} {Phys. Rev. A}\ }\textbf {\bibinfo {volume} {78}},\ \bibinfo
  {pages} {032316} (\bibinfo {year} {2008})}\BibitemShut {NoStop}%
\bibitem [{\citenamefont {Mari}\ and\ \citenamefont {Eisert}(2009)}]{Mari2009}%
  \BibitemOpen
  \bibfield  {author} {\bibinfo {author} {\bibfnamefont {A.}~\bibnamefont
  {Mari}}\ and\ \bibinfo {author} {\bibfnamefont {J.}~\bibnamefont {Eisert}},\
  }\bibfield  {title} {\bibinfo {title} {{``Gently Modulating Optomechanical
  Systems''}},\ }\href {https://doi.org/10.1103/PhysRevLett.103.213603}
  {\bibfield  {journal} {\bibinfo  {journal} {Phys. Rev. Lett.}\ }\textbf
  {\bibinfo {volume} {103}},\ \bibinfo {pages} {213603} (\bibinfo {year}
  {2009})}\BibitemShut {NoStop}%
\bibitem [{\citenamefont {Plenio}(2005)}]{Plenio2005}%
  \BibitemOpen
  \bibfield  {author} {\bibinfo {author} {\bibfnamefont {M.~B.}\ \bibnamefont
  {Plenio}},\ }\bibfield  {title} {\bibinfo {title} {{``Logarithmic Negativity:
  A Full Entanglement Monotone That is not Convex''}},\ }\href
  {https://doi.org/10.1103/PhysRevLett.95.090503} {\bibfield  {journal}
  {\bibinfo  {journal} {Phys. Rev. Lett.}\ }\textbf {\bibinfo {volume} {95}},\
  \bibinfo {pages} {090503} (\bibinfo {year} {2005})}\BibitemShut {NoStop}%
\bibitem [{\citenamefont {Rabl}(2011)}]{Rabl2011}%
  \BibitemOpen
  \bibfield  {author} {\bibinfo {author} {\bibfnamefont {P.}~\bibnamefont
  {Rabl}},\ }\bibfield  {title} {\bibinfo {title} {{``Photon Blockade Effect in
  Optomechanical Systems''}},\ }\href
  {https://doi.org/10.1103/PhysRevLett.107.063601} {\bibfield  {journal}
  {\bibinfo  {journal} {Phys. Rev. Lett.}\ }\textbf {\bibinfo {volume} {107}},\
  \bibinfo {pages} {063601} (\bibinfo {year} {2011})}\BibitemShut {NoStop}%
\bibitem [{\citenamefont {Carmichael}(1999)}]{Carmichael1999}%
  \BibitemOpen
  \bibfield  {author} {\bibinfo {author} {\bibfnamefont {H.~J.}\ \bibnamefont
  {Carmichael}},\ }\href {https://doi.org/10.1007/978-3-662-03875-8} {\emph
  {\bibinfo {title} {{``Statistical Methods in Quantum Optics 1''}}}}\
  (\bibinfo  {publisher} {Springer-Verlag},\ \bibinfo {address} {Berlin},\
  \bibinfo {year} {1999})\BibitemShut {NoStop}%
\bibitem [{\citenamefont {Jaehne}\ \emph {et~al.}(2008)\citenamefont {Jaehne},
  \citenamefont {Hammerer},\ and\ \citenamefont {Wallquist}}]{Jaehne2008}%
  \BibitemOpen
  \bibfield  {author} {\bibinfo {author} {\bibfnamefont {K.}~\bibnamefont
  {Jaehne}}, \bibinfo {author} {\bibfnamefont {K.}~\bibnamefont {Hammerer}},\
  and\ \bibinfo {author} {\bibfnamefont {M.}~\bibnamefont {Wallquist}},\
  }\bibfield  {title} {\bibinfo {title} {{``Ground-state cooling of a
  nanomechanical resonator via a Cooper-pair box qubit''}},\ }\href
  {https://doi.org/10.1088/1367-2630/10/9/095019} {\bibfield  {journal}
  {\bibinfo  {journal} {New J. Phys.}\ }\textbf {\bibinfo {volume} {10}},\
  \bibinfo {pages} {095019} (\bibinfo {year} {2008})}\BibitemShut {NoStop}%
\bibitem [{\citenamefont {Johansson}\ \emph {et~al.}(2012)\citenamefont
  {Johansson}, \citenamefont {Nation},\ and\ \citenamefont
  {Nori}}]{Johansson2012}%
  \BibitemOpen
  \bibfield  {author} {\bibinfo {author} {\bibfnamefont {J.~R.}\ \bibnamefont
  {Johansson}}, \bibinfo {author} {\bibfnamefont {P.~D.}\ \bibnamefont
  {Nation}},\ and\ \bibinfo {author} {\bibfnamefont {F.}~\bibnamefont {Nori}},\
  }\bibfield  {title} {\bibinfo {title} {{``QuTiP: An open-source Python
  framework for the dynamics of open quantum systems''}},\ }\href
  {https://doi.org/10.1016/j.cpc.2012.02.021} {\bibfield  {journal} {\bibinfo
  {journal} {Comp. Phys. Comm.}\ }\textbf {\bibinfo {volume} {183}},\ \bibinfo
  {pages} {1760} (\bibinfo {year} {2012})}\BibitemShut {NoStop}%
\bibitem [{\citenamefont {Adesso}\ \emph {et~al.}(2014)\citenamefont {Adesso},
  \citenamefont {Ragy},\ and\ \citenamefont {Lee}}]{Adesso2014}%
  \BibitemOpen
  \bibfield  {author} {\bibinfo {author} {\bibfnamefont {G.}~\bibnamefont
  {Adesso}}, \bibinfo {author} {\bibfnamefont {S.}~\bibnamefont {Ragy}},\ and\
  \bibinfo {author} {\bibfnamefont {A.~R.}\ \bibnamefont {Lee}},\ }\bibfield
  {title} {\bibinfo {title} {{``Continuous Variable Quantum Information:
  Gaussian States and Beyond''}},\ }\href
  {https://doi.org/10.1142/S1230161214400010} {\bibfield  {journal} {\bibinfo
  {journal} {Open Syst. Inf. Dyn.}\ }\textbf {\bibinfo {volume} {21}},\
  \bibinfo {pages} {1440001} (\bibinfo {year} {2014})}\BibitemShut {NoStop}%
\bibitem [{\citenamefont {Krischek}\ \emph {et~al.}(2011)\citenamefont
  {Krischek}, \citenamefont {Schwemmer}, \citenamefont {Wieczorek},
  \citenamefont {Weinfurter}, \citenamefont {Hyllus}, \citenamefont
  {Pezz{\'e}},\ and\ \citenamefont {Smerzi}}]{Krischek2011}%
  \BibitemOpen
  \bibfield  {author} {\bibinfo {author} {\bibfnamefont {R.}~\bibnamefont
  {Krischek}}, \bibinfo {author} {\bibfnamefont {C.}~\bibnamefont {Schwemmer}},
  \bibinfo {author} {\bibfnamefont {W.}~\bibnamefont {Wieczorek}}, \bibinfo
  {author} {\bibfnamefont {H.}~\bibnamefont {Weinfurter}}, \bibinfo {author}
  {\bibfnamefont {P.}~\bibnamefont {Hyllus}}, \bibinfo {author} {\bibfnamefont
  {L.}~\bibnamefont {Pezz{\'e}}},\ and\ \bibinfo {author} {\bibfnamefont
  {A.}~\bibnamefont {Smerzi}},\ }\bibfield  {title} {\bibinfo {title}
  {{``Useful Multiparticle Entanglement and Sub-Shot-Noise Sensitivity in
  Experimental Phase Estimation''}},\ }\href
  {https://doi.org/10.1103/PhysRevLett.107.080504} {\bibfield  {journal}
  {\bibinfo  {journal} {Phys. Rev. Lett.}\ }\textbf {\bibinfo {volume} {107}},\
  \bibinfo {pages} {080504} (\bibinfo {year} {2011})}\BibitemShut {NoStop}%
\bibitem [{\citenamefont {Hyllus}\ \emph {et~al.}(2012)\citenamefont {Hyllus},
  \citenamefont {Laskowski}, \citenamefont {Krischek}, \citenamefont
  {Schwemmer}, \citenamefont {Wieczorek}, \citenamefont {Weinfurter},
  \citenamefont {Pezze},\ and\ \citenamefont {Smerzi}}]{Hyllus2012}%
  \BibitemOpen
  \bibfield  {author} {\bibinfo {author} {\bibfnamefont {P.}~\bibnamefont
  {Hyllus}}, \bibinfo {author} {\bibfnamefont {W.}~\bibnamefont {Laskowski}},
  \bibinfo {author} {\bibfnamefont {R.}~\bibnamefont {Krischek}}, \bibinfo
  {author} {\bibfnamefont {C.}~\bibnamefont {Schwemmer}}, \bibinfo {author}
  {\bibfnamefont {W.}~\bibnamefont {Wieczorek}}, \bibinfo {author}
  {\bibfnamefont {H.}~\bibnamefont {Weinfurter}}, \bibinfo {author}
  {\bibfnamefont {L.}~\bibnamefont {Pezze}},\ and\ \bibinfo {author}
  {\bibfnamefont {A.}~\bibnamefont {Smerzi}},\ }\bibfield  {title} {\bibinfo
  {title} {{``Fisher information and multiparticle entanglement''}},\ }\href
  {https://doi.org/10.1103/PhysRevA.85.022321} {\bibfield  {journal} {\bibinfo
  {journal} {Phys. Rev. A}\ }\textbf {\bibinfo {volume} {85}},\ \bibinfo
  {pages} {022321} (\bibinfo {year} {2012})}\BibitemShut {NoStop}%
\bibitem [{\citenamefont {T{\'o}th}(2012)}]{Toth2012}%
  \BibitemOpen
  \bibfield  {author} {\bibinfo {author} {\bibfnamefont {G.}~\bibnamefont
  {T{\'o}th}},\ }\bibfield  {title} {\bibinfo {title} {{``Multipartite
  entanglement and high-precision metrology''}},\ }\href
  {https://doi.org/10.1103/PhysRevA.85.022322} {\bibfield  {journal} {\bibinfo
  {journal} {Phys. Rev. A}\ }\textbf {\bibinfo {volume} {85}},\ \bibinfo
  {pages} {022322} (\bibinfo {year} {2012})}\BibitemShut {NoStop}%
\bibitem [{\citenamefont {Genoni}\ and\ \citenamefont
  {Paris}(2010)}]{Genoni2010}%
  \BibitemOpen
  \bibfield  {author} {\bibinfo {author} {\bibfnamefont {M.~G.}\ \bibnamefont
  {Genoni}}\ and\ \bibinfo {author} {\bibfnamefont {M.~G.~A.}\ \bibnamefont
  {Paris}},\ }\bibfield  {title} {\bibinfo {title} {{``Quantifying
  non-Gaussianity for quantum information''}},\ }\href
  {https://doi.org/10.1103/PhysRevA.82.052341} {\bibfield  {journal} {\bibinfo
  {journal} {Phys. Rev. A}\ }\textbf {\bibinfo {volume} {82}},\ \bibinfo
  {pages} {052341} (\bibinfo {year} {2010})}\BibitemShut {NoStop}%
\end{thebibliography}%

\end{document}